\documentclass[conference]{IEEEtran}
\IEEEoverridecommandlockouts
\usepackage{cite}
\usepackage{amsmath,amssymb,amsfonts}
\usepackage{graphicx}
\usepackage{textcomp}
\usepackage{xcolor}
\usepackage{caption}
\usepackage{subcaption}
\def\BibTeX{{\rm B\kern-.05em{\sc i\kern-.025em b}\kern-.08em
    T\kern-.1667em\lower.7ex\hbox{E}\kern-.125emX}}
\usepackage{booktabs}
\usepackage[ruled, vlined, linesnumbered]{algorithm2e}
\usepackage{algorithmic}
\usepackage{diagbox}
\usepackage{balance}
\usepackage{multirow}
\usepackage{epstopdf}
\usepackage{float}
\usepackage{url}
\usepackage[hidelinks]{hyperref}
\usepackage{enumerate}
\usepackage{afterpage}
\usepackage{setspace}

\newtheorem{definition}{Definition} 
\newtheorem{example}{Example} 

\newtheorem{lemma}{Lemma} 
\newtheorem{theorem}{Theorem} 

\newcommand{\ie}{\emph{i.e.},\xspace}
\newcommand{\eg}{\emph{e.g.},\xspace}
\newcommand{\st}{\emph{s.t.}\xspace}
\newcommand{\aka}{\emph{a.k.a.},\xspace}

\newcommand{\dist}{{\mathsf{d}}}
\newcommand{\cord}[1]{\langle {#1} \rangle}

\newcommand\figref[1]{Fig.~\ref{#1}}

\newcommand\tabref[1]{Table~\ref{#1}}
\newcommand\algoref[1]{Algorithm~\ref{#1}}

\newcommand\secref[1]{Sec.~\ref{#1}}
\newcommand\defref[1]{Definition~\ref{#1}}
\newcommand\equref[1]{Equation~(\ref{#1})}

\newcommand\algref[1]{Algorithm~\ref{#1}}
\newcommand\lemref[1]{Lemma~\ref{#1}}
\newcommand\thmref[1]{Theorem~\ref{#1}}

\newcommand\continue{\textbf{continue}\;}
\newcommand\framework{\underline{G}eo-\underline{I} accelerated \underline{S}MC based method for federated \underline{T}rajectory matching\xspace}
\newcommand\frameworkabbr{{\textsf{GIST}}\xspace}
\newcommand\problem{{Federated Trajectory Matching}\xspace}
\newcommand\problemabbr{FTM\xspace}
\newcommand\match{\textsf{match}}

\newcommand{\fakeparagraph}[1]{\vspace{1mm}\noindent\textbf{#1.}}

\ifodd 0
\newcommand{\TODO}[1]{\textbf{\color{red}{TODO: #1} }}

\else

\newcommand{\TODO}[1]{}

\fi

\begin{document}
\bstctlcite{IEEEexample:BSTcontrol}

\title{
Efficient and Private Federated Trajectory Matching
}


\author{
\IEEEauthorblockN{Yuxiang Wang$^{\dagger}$, Yuxiang Zeng$^{\dagger}$, Yi Xu$^{\dagger}$, Zimu Zhou$^{\ddagger}$, Yongxin Tong$^{\dagger}$}
\IEEEauthorblockA{
$^{\dagger}$ SKLSDE Lab, BDBC and IRI, Beihang University, Beijing, China \\
$^{\ddagger}$ City University of Hong Kong, Hong Kong, China \\
$^{\dagger}$\{yuxiangwang, yxzeng, xuy, yxtong\}@buaa.edu.cn, $^{\ddagger}$zimuzhou@cityu.edu.hk}
}

\maketitle
\thispagestyle{plain}

\begin{abstract}
Federated Trajectory Matching (\problemabbr) is gaining increasing importance in big trajectory data analytics, supporting diverse applications such as public health, law enforcement, and emergency response.
\problemabbr retrieves trajectories that match with a query trajectory from a large-scale trajectory database, while safeguarding the privacy of trajectories in both the query and the database. 
A naive solution to \problemabbr is to process the query through Secure Multi-Party Computation (SMC) across the entire database, which is inherently secure yet inevitably slow due to the massive secure operations.
A promising acceleration strategy is to filter irrelevant trajectories from the database based on the query, thus reducing the SMC operations. 
However, a key challenge is how to publish the query in a way that both preserves privacy and enables efficient trajectory filtering. 
In this paper, we design \frameworkabbr, a novel framework for efficient Federated Trajectory Matching. 
\frameworkabbr is grounded in Geo-Indistinguishability, a privacy criterion dedicated to locations.
It employs a new privacy mechanism for the query that facilitates efficient trajectory filtering.
We theoretically prove the privacy guarantee of the mechanism and the accuracy of the filtering strategy of \frameworkabbr.
Extensive evaluations on five real datasets show that \frameworkabbr is significantly faster and incurs up to 3 orders of magnitude lower communication cost than the state-of-the-arts.
\end{abstract}


\section{Introduction}
The emergence of big trajectory data, powered by diverse sensors such as GPS, surveillance cameras, and proximity sensors, has revolutionized our ability to capture and analyze movement patterns \cite{zheng2015trajectory}.
This data, often collected by various entities ranging from tech companies to government agencies, offers a multifaceted view of human mobility and urban activities. 
However, the distributed nature of data ownership, coupled with the inherent sensitivity of trajectory data \cite{chow2011trajectory,de2013unique}, necessitates paradigms that respect privacy constraints while enabling effective analysis across different data owners.

Of our particular interest is Federated Trajectory Matching (\problemabbr), a primitive in privacy-preserving trajectory analysis across distributed data owners. 
\problemabbr retrieves trajectories in a large-scale private dataset, held by a distinct data owner, that match with a query trajectory. 
Importantly, this query process should safeguard two categories of trajectory privacy:
\textit{(i)} the exact spatiotemporal information in the query trajectory; and \textit{(ii)} any trajectories in the database other than the query result. 
We illustrate the use cases of \problem query via the following real-world applications.

\begin{example}[Tracing Infections in Epidemics \cite{epidemic}] \label{epidemic}
During a contagious disease outbreak, health officials often face the task of tracing infection paths from a limited location history.
They may turn to the trajectory database of the Location-Based Service (LBS) providers. 
However, the raw location history is confidential, as its disclosure might induce panic. 
Likewise, it is crucial for the LBS provider to prevent the trace of the uninfected individual from leakage.
\end{example}

\begin{example}[Tracking Criminal Suspects \cite{li2021trajectory,criminal}]\label{exp:criminal}
The police often locate a criminal suspect by analyzing trajectory data from surveillance cameras or witnesses.
They can improve the tracking by collaborating with LBS providers via the dense GPS trajectories.
However, regulations strictly limit the sharing of sensitive trajectory data with law enforcement \cite{GDPR,CCPA}, and the police are equally constrained from providing the raw query trajectory to LBS providers, as these may contain confidential information. 
\end{example}

A central challenge in \problemabbr is to attain high query efficiency over large-scale data. 
While Secure Multi-Party Computation (SMC) effectively ensures privacy, it falls short in terms of efficiency. 
As our empirical study (\secref{sec:exp-real}) shows, processing a single \problemabbr query on a database containing 3.2 million trajectories using generic SMC techniques \cite{bayatbabolghani2018secure,lindell2020secure} or those specialized for trajectory similarity \cite{liu2015efficient} can take as long as 89 hours. 
Such processing times are impractical in situations where swift responses are critical, such as in managing public health emergencies or conducting criminal investigations. 
This inefficiency arises from the necessity to process \textit{every} trajectory in the database via SMC operations, a requirement that significantly hampers the feasibility of these methods in time-sensitive and large-scale applications.

Observing that typically less than 1\% trajectories in the database matches the query, 
we propose a simple acceleration strategy: \textit{filtering} trajectories unlikely to match the query to reduce the number of SMC operations, while simultaneously maintaining privacy.
Realizing this strategy, however, is non-trivial.
The two main challenges are: 
\textit{(i)} designing a privacy mechanism that enables accurate trajectory filtering, and \textit{(ii)} developing an effective filtering scheme that operates on perturbed query trajectories.
Previous studies mainly focused on privacy mechanisms and acceleration techniques for secure queries in relational databases \cite{wagh2021dp, he2017composing, bater2018shrinkwrap, wang2021dp, wang2022incshrink}, which do not easily translate to trajectory matching due to inherent differences in data structures and operations.
Furthermore, conventional privacy mechanisms for location or trajectory data \cite{andres2013geo, cunningham2021real, zhang2023trajectory} are not optimized for trajectory filtering, leading to high retention rate (see \secref{sec:exp-ablation}).

To this end, we present \frameworkabbr, an efficient framework for \problemabbr queries. 
\frameworkabbr is grounded in Geo-Indistinguishability (Geo-I) \cite{andres2013geo}, a recognized differential privacy standard for location data. 
It incorporates novel privacy mechanisms and trajectory filtering strategies tailored to \problemabbr. 
Specifically, the query trajectory is perturbed using a newly devised Bounded Planar Laplace (BPL) mechanism and then shared with the data owner at a grid level, allowing the data owner to conduct effective trajectory filtering. 
The trade-off between the trajetory filtering granularity and the privacy parameters is analyzed theoretically. 
Moreover, we devise a data partition scheme along with a reference trajectory based pruning strategy to further improve the query efficiency.

Our major contributions are summarized as follows:
\begin{itemize}
    \item 
    We define \problem (\problemabbr), an emerging problem in privacy-aware big trajectory data analysis that has various real-world applications.
    \item 
    We propose \frameworkabbr, a framework to accelerate FTM on large-scale data while accounting for privacy.
    The key is to reduce the number of secure operations for trajectory matching via Geo-Indistinguishability.
    To the best of our knowledge, this is the first work that adopts this strategy to trajectory matching.
    \item 
    We develop a novel grid-level query trajectory publishing method which ensures both privacy guarantee and trajectory filtering efficiency.
    We theoretically analyze the trade-off between privacy level and filtering efficiency.
    \item 
    Extensive experiments on five real datasets show that our solution outperforms the state-of-the-arts \cite{oblivc,liu2015efficient} by a large margin. 
\end{itemize}

The rest of paper is organized as follows. 
In \secref{sec:pre}, we present the problem definition and related concepts. 
Then, we introduce the overall framework in \secref{sec:overview} and elaborate on the technical details in \secref{sec:alg}. 
Finally, we conduct the experimental evaluation in \secref{sec:exp}, review existing studies in \secref{sec:related}, and conclude in \secref{sec:conclusion}.
\section{Preliminaries} \label{sec:pre}
This section presents the problem definition (\secref{sec:pre-pro}) and some prerequisites on Geo-Indistinguishability (\secref{sec:pre-geoi}).

\subsection{Problem Definition} \label{sec:pre-pro}

\begin{definition}[Point \cite{wang2021survey}]
Each point $p$ is denoted by a timestamp $p.ts$ and the geo-location $p.loc$ at this timestamp.
\end{definition}

For any two points $p,q$, the Euclidean distance function $\dist(p,q)$ computes the distance between $p$ and $q$.

\begin{definition}[Trajectory \cite{wang2021survey}]
    A trajectory $T$ is defined as a sequence of $|T|$ points, \ie $T=\{p_1,p_2,\ldots,p_{|T|}\}$.
\end{definition}

In practice, points in a trajectory can be simplified as a piecewise linear function of the timestamp \cite{vsaltenis2000indexing} and each piece of the function is defined as a segment in the following.

\begin{definition}[Segment \cite{vsaltenis2000indexing}]
    A segment $s=\{o,d\}$ is represented by a pair of points. The points $o$ and $d$ represent the origin and destination points of the segment, and satisfy the timestamp condition $o.ts \le d.ts$.
\end{definition}

Linear interpolation is usually employed to derive the location of a segment $s$ at any timestamp \cite{vsaltenis2000indexing}. Specifically, we calculate the velocity of the segment as $\overline{v}=\frac{d.loc-o.loc}{d.ts-o.ts}$ and estimate the location of $s$ at timestamp $ts'$ as $loc_s(ts')=o.loc+(ts'-o.ts)\cdot \overline{v}$. In addition, the location of a trajectory $T$ at timestamp $ts'$ is computed as $loc_T(ts')=loc_s(ts')$, where $s$ is a segment in $T$ and satisfies $ts'\in [s.o.ts,s.d.ts]$.

\begin{figure}[htbp]
\centerline{\includegraphics[width=0.3\textwidth]{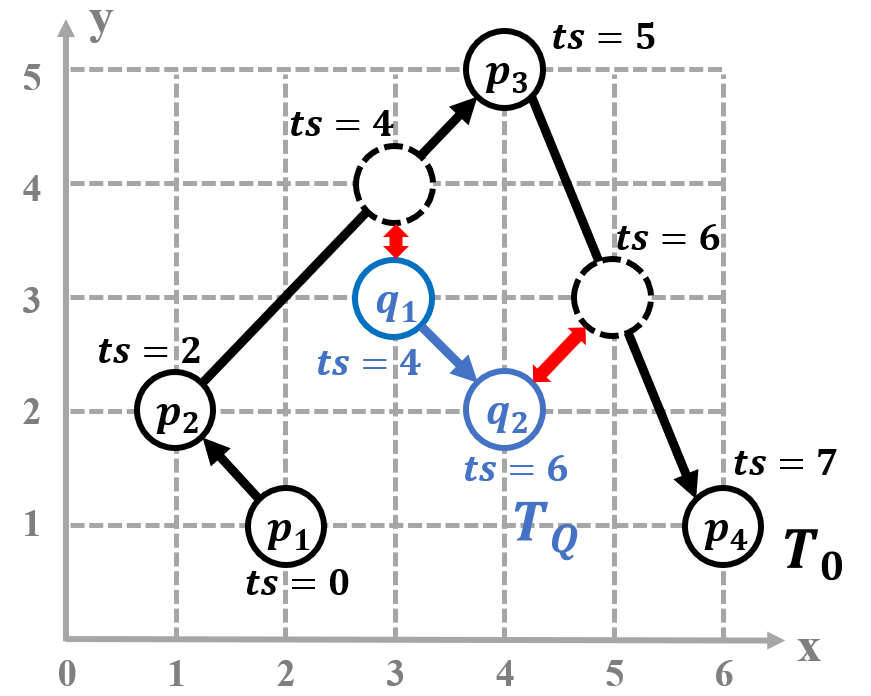}}
\caption{The example of trajectory match.}
\label{fig:match}
\end{figure}

\begin{example}\label{exp:traj}
    Consider trajectory $T_0=\{p_1,p_2,p_3,p_4\}$ in \figref{fig:match}, where $p_1.loc=\cord{2,1}$, $p_1.ts=0$, $p_2.loc=\cord{1,2}$, $p_2.ts=2$, $p_3.loc=\cord{4,5}$, $p_3.ts=5$, $p_4.loc=\cord{6,1}$, $p_4.ts=7$. Trajectory $T_0$ can be seen as a sequence of segments $\{s_1,s_2,s_3\}$, where $s_1=\{p_1,p_2\}$, $s_2=\{p_2,p_3\}$, $s_3=\{p_3,p_4\}$. We apply linear interpolation to segment $s_2$ to estimate the location of $T_0$ at timestamp $4$. Namely, $\overline{v}=\frac {(\cord{4,5}-\cord{1,2})} {5-2}=\cord{1,1}$, $loc_{T_0}(4)=\cord{1,2}+(4-2)\cdot\overline{v}=\cord{3,4}$.
\end{example}

\begin{definition}[Trajectory Matching]\label{def:match}
Given a distance threshold $\tau$ and a trajectory $T_Q$, a trajectory $T_i$ is considered to be matched with $T_Q$, denoted as $\match_{\tau}(T_i,T_Q)=\textsf{true}$, if for every point $q\in T_Q$:
\begin{equation}
    \dist(q.loc,\ loc_{T_i}(q.ts))\le\tau
\end{equation}
where $loc_{T_i}(q.ts)$ represents the location of trajectory $T_i$ with the same timestamp as point $q$.
\end{definition}

In practice, trajectory matching requires that each location in $T_Q$ has a corresponding location in $T_i$ that is sufficiently close (\ie $\le \tau$). 
The definition is akin to the frequently-used spatiotemporal distance measure STED\cite{nanni2006time,su2020survey}, and the requirement that each location in the query trajectory be matched makes it more suitable for our application scenario.

\begin{example}\label{exp:traj-match}
Consider trajectory $T_0$ and $T_Q$ in \figref{fig:match}. The query trajectory $T_Q=\{q_1,q_2\}$, where $q_1.loc=\cord{3,3},q_1.ts=4$, $q_2.loc=\cord{4,2},q_2.ts=6$. We set the distance threshold $\tau=1.5$. According to the definition, we examine timestamps $4$ and $6$, computing $loc_{T_0}(4)=\cord{3,4}$ and $loc_{T_0}(6)=\cord{5,3}$. Because $\dist(q_1.loc,loc_{T_0}(4))=\sqrt{0^2+1^2}=1 < 1.5$ and $\dist(q_2.loc,loc_{T_0}(6))=\sqrt{1^2+1^2}= \sqrt 2 < 1.5$, we conclude that $T_0$ can be matched with $T_Q$.
\end{example}

\begin{definition}[Trajectory Data Federation]
The trajectory data federation comprises one or more data owners, each autonomously managing their local trajectory data. When a user submits a query, data owners and the query user collaborate to execute the queries \cite{bater2017smcql,tong2022hu}. Due to the high sensitivity of trajectory data \cite{chow2011trajectory,de2013unique}, trajectories other than the query result cannot be leaked to the query user or other data owners during the query execution.
\end{definition}

\begin{definition}[\problem (\problemabbr)]
Given a trajectory data federation $TD$ containing a large amount of trajectories, a query trajectory $T_Q$, and a distance threshold $\tau$, the query $FTM(TD,T_Q)$ aims to securely retrieve all trajectories in $TD$ that match with $T_Q$:
\begin{equation*}
    FTM(TD,T_Q)=\{T_i|T_i\in TD \wedge \match_{\tau}(T_i, T_Q)=\textsf{true}\}
\end{equation*}
It is required that the \problemabbr query procedure prevents the leakage of spatiotemporal information about points in $T_Q$ to the data owner. Besides, information about unmatched trajectories in $TD$ cannot be disclosed to the query user.
\end{definition}

\subsection{Geo-Indistinguishability for Protecting Location Privacy} \label{sec:pre-geoi}

Geo-Indistinguishability (Geo-I) \cite{andres2013geo} extends the de facto standard notion of privacy protection, \ie $\epsilon$-differential privacy ($\epsilon$-DP) \cite{dwork2006calibrating}, to spatial data. Geo-I is widely adopted in the location-based systems and can be utilized to safeguard privacy in \problemabbr.
A mechanism $M$ operates as a probabilistic function, taking any location within $\mathbb{X}$ as input and mapping it into a location within $\mathbb{Y}$ as output.

\begin{definition}[Geo-Indistinguishability ($\epsilon$-Geo-I) \cite{andres2013geo}]
A mechanism $M$ satisfies $\epsilon$-Geo-Indistinguishability ($\epsilon$-Geo-I) iff for all $x, x'\in \mathbb{X}$ and all $Y \subseteq \mathbb{Y}$:
\begin{equation}
    Pr[M(x)\in Y]\le e^{\epsilon \dist(x,x')}Pr[M(x')\in Y]
\end{equation}
\end{definition}

\fakeparagraph{Planar Laplace Mechanism} 
Geo-Indistinguishability is usually achieved by introducing planar Laplacian noise \cite{andres2013geo}, which can be generated through independent samples of the radial distance $r$ and polar angle $\theta$ in the plane polar coordinates.

The radius $r$ depends on the cumulative distribution function (CDF) $C_{\epsilon}(r)$:
\begin{equation} \label{eq:cdf}
    C_{\epsilon} (r)=1-(1+\epsilon r)e^{-\epsilon r} 
\end{equation}

To derive the radius $r$ from a given probability $p$, we can use the inverse function of $p=C_{\epsilon}(r)$, denoted as $C_{\epsilon}^{-1} (p)$:

\begin{equation} \label{eq:cdf-rev}
C_{\epsilon}^{-1} (p)=-\frac{1}{\epsilon} \left[W_{-1}\left(\frac{p-1}{e}\right) + 1\right]
\end{equation}
where $W_{-1}$ represents the Lambert W function's $-1$ branch.

When generating the planar Laplacian noise, we first randomly pick $p$ from the uniform distribution within $[0,1]$ and then obtain $r=C_{\epsilon}^{-1}(p)$. After that, we choose $\theta \in [0,2\pi]$ uniformly at random, and compute the noise as $\cord{rcos \theta, r sin \theta}$.

\section{Framework Overview} \label{sec:overview}

\begin{figure}[htbp]
\centerline{\includegraphics[width=0.48\textwidth]{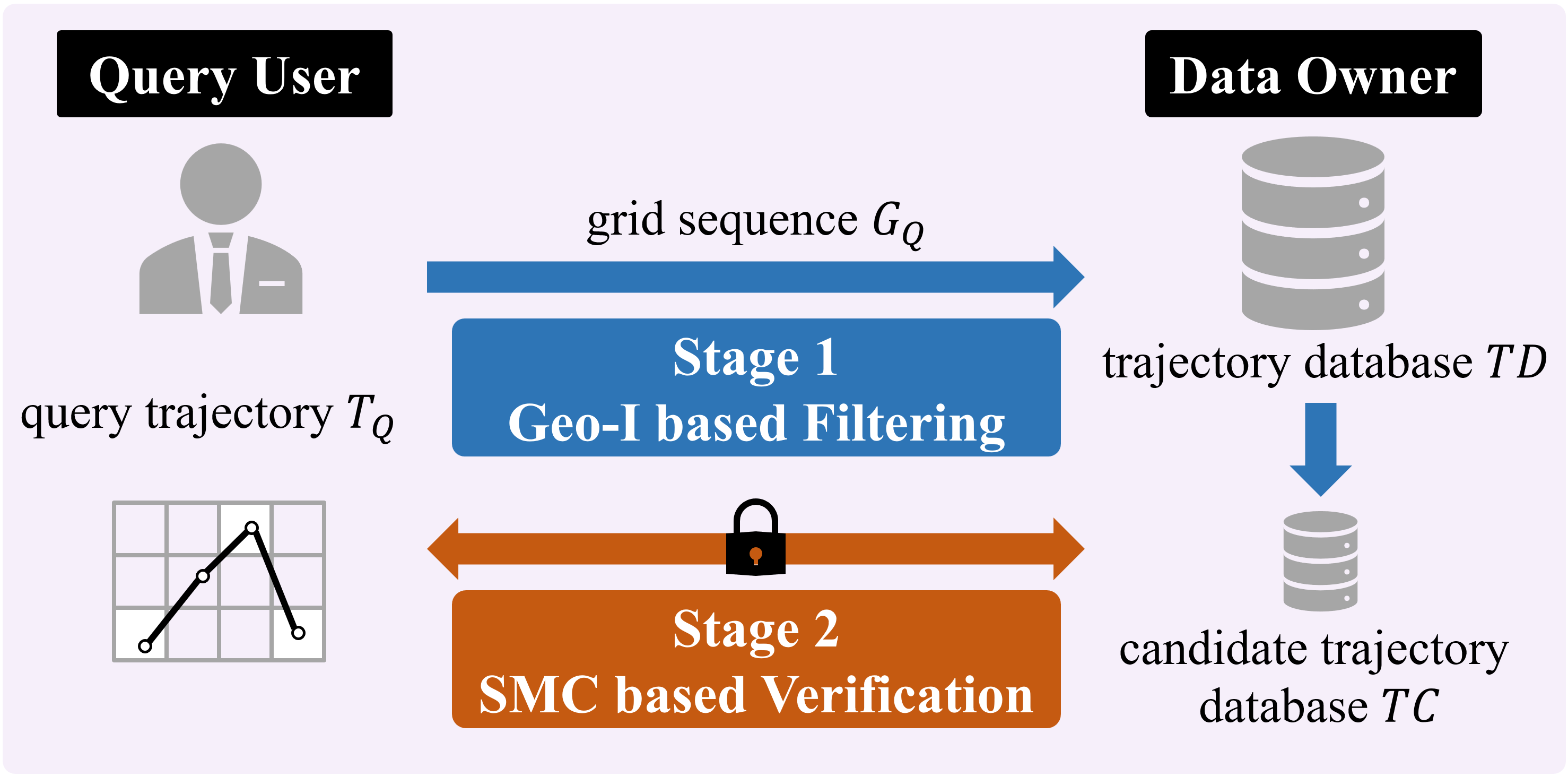}}
\caption{\framework (\frameworkabbr).}
\label{fig:framework}
\end{figure}

To alleviate the high time cost of SMC in \problemabbr query, we design a novel framework called \textbf{\framework (\frameworkabbr)}.
Utilizing a filtering strategy that adheres to the constraint of Geo-Indistinguishability, we circumvent the need for scanning the entire trajectory database via SMC operations and accelerate the \problemabbr query significantly.
\frameworkabbr comprises the following two phases:

\begin{itemize}
    \item \textbf{Geo-I based Filtering:} Initially, the user processes the query trajectory $T_Q$ and publish it at a grid level ($G_Q$ in \figref{fig:framework}). The procedure of trajectory publishing complies with the privacy constraint of $(\epsilon,\delta)$-Geo-I, a relaxation of standard Geo-I. Subsequently, the data owner utilizes $G_Q$ to locally filter the database $TD$ and obtain a reduced database $TC$, with the grid index accelerating the computation. 
    \item \textbf{SMC based Verification:} Following the filtering phase, both the query user and the data owner securely verify trajectories within $TC$ to identify all trajectories that match $T_Q$. We devise a data partition scheme along with a reference trajectory based pruning strategy to further improve efficiency. 
\end{itemize}

We focus on the scenario with a single data owner, since the \problemabbr in a data federation with multiple data owners can be addressed by executing the FTM query with each data owner in parallel (see \secref{sec:exp-fed}). The major notations used in the paper are listed in \tabref{tab:symbol}.

\begin{table}[t]\footnotesize
\centering
\caption{Summary of major notations.}\label{tab:symbol}
\begin{tabular}{|c|c|} \hline
    Notations & Description \\ \hline
    $TD,TC$ & trajectory database and candidate trajectory database \\ \hline
    $T_Q,\tau$ & query trajectory and distance threshold \\ \hline
    $\epsilon,\delta$ & privacy budget and failure probability\\ \hline
    $G_Q,GI$ & published grid sequence and grid index\\ \hline
    $L,R$ & grid size and radius of noise circle \\ \hline
    $p$ & probability of successfully perturbing a single location \\ \hline
    $PN,rt$ & partition and its reference trajectory \\ \hline
    $\alpha, m$ & partition parameter and maximum size of partition \\ \hline
\end{tabular}
\end{table}
\section{Algorithm Design} \label{sec:alg}
This section introduces the algorithm designs of our framework \frameworkabbr from two aspects:
\textit{Geo-I based Filtering} (\secref{sec:alg-filter}) and \textit{SMC based Verification} (\secref{sec:alg-verify}).

\subsection{Geo-I based Filtering} \label{sec:alg-filter}
In the following, we first introduce a location privacy definition named $(\epsilon,\delta)$-Geo-I and propose a mechanism to achieve it.
Then, we provide a detailed explanation of how to filter candidate answers from $TD$ based on the privately published query trajectory.
Finally, we theoretically derive the appropriate grid size used in the filtering process.

\subsubsection{\textbf{$(\epsilon,\delta)$-Geo-Indistinguishability and Bounded Planar Laplace Mechanism}} \label{sec:alg-filter-geoi}

To pursue a promising query performance, we propose a new definition of location privacy based on $\epsilon$-Geo-I and achieve it with a mechanism named Bounded Planar Laplace (BPL).

\fakeparagraph{Motivation} 
In $\epsilon$-Geo-I, the spatial range of the injected noise is usually unbounded under Planar Laplace mechanism, meaning that the original location can be perturbed to an arbitrarily distant location.
In practice, this feature may lead to unexpected result: a perturbed location too far away from the original one may seriously compromise the usability (\ie query performance in our case).
Thus, we aim to design a new privacy mechanism, which can not only restrict the upper bound of noise in the spatial area, but also generally follows the concept of $\epsilon$-Geo-I.

\fakeparagraph{Definition of $(\epsilon,\delta)$-Geo-I}
Motivated by the generalization of $(\epsilon,\delta)$-DP (\aka approximate DP) from $\epsilon$-DP (\aka pure DP) \cite{dwork2014algorithmic},
we introduce an approximate version of Geo-I in \defref{def:delta-Geo-I}, allowing a small probability $\delta$ of failing to reach $\epsilon$-Geo-I.

\begin{definition}[$(\epsilon,\delta)$-Geo-I]\label{def:delta-Geo-I}
A mechanism $M$ satisfies $(\epsilon,\delta)$-Geo-Indistinguishability ($(\epsilon,\delta)$-Geo-I) iff for all $x, x'\in \mathbb{X}$ and all $Y \subseteq \mathbb{Y}$:
\begin{equation}
    Pr[M(x)\in Y]\le e^{\epsilon \dist(x,x')}Pr[M(x')\in Y]+\delta
\end{equation}
\end{definition}

Post-processing is a crucial property for differential privacy \cite{dwork2014algorithmic}. Similarly, we can prove in \lemref{lem:post-process} that this property holds true for $(\epsilon,\delta)$-Geo-I as well.

\begin{lemma}[Post-processing]\label{lem:post-process}
Give mechanism $M$ that satisfies $(\epsilon,\delta)$-Geo-I, then for any algorithm $f$, the composition of $M$ and $f$, \ie $f(M(\cdot))$ satisfies $(\epsilon,\delta)$-Geo-I.
\end{lemma}
\begin{IEEEproof}
We first prove the result for any deterministic function $f$.
The lemma then follows as any randomized mapping can be decomposed into a convex combination of deterministic functions \cite{dwork2014algorithmic}.
Define the output domain of $f$ as $\mathbb{Z}$. For any $x,x'\in \mathbb{X}$, and any $Z \subseteq \mathbb{Z}$, we prove this lemma as follows:
\begin{align*}
Pr[f(M(x))\in Z]
&=Pr[M(x)\in Y]\\
&\le e^{\epsilon \dist(x,x')}Pr[M(x')\in Y]+\delta\\
&=e^{\epsilon \dist(x,x')}Pr[f(M(x'))\in Z]+\delta
\end{align*}
where $Y=\{y\in \mathbb{Y}|f(y)\in Z\}$.
\end{IEEEproof}

\fakeparagraph{Privacy Mechanism: Bounded Planar Laplace}
We devise the Bounded Planar Laplace (BPL) mechanism to achieve $(\epsilon,\delta)$-Geo-I.
The \textit{main advantage} of the BPL mechanism lies in its ability to constrain the maximum value of noise.
In other words, the distance between the perturbed location and the original location cannot exceed $R$.
For conciseness, we refer to the circle with a radius of $R$ as the \textit{noise circle}.

\begin{algorithm}[t]
\SetKwInOut{Input}{input}
\SetKwInOut{Output}{output}
\Input{location $x$, privacy parameters $\epsilon,\delta$}
\Output{perturbed location $x'$}
Find $\Delta$ that satisfies $\Delta=\delta \pi[C_{\epsilon}^{-1} (1-\Delta)] ^2$\;
$R\gets C_{\epsilon}^{-1} (1-\Delta)$\;
Choose $p \in [0,1]$ uniformly at random\;
\If(\tcp*[f]{planar Laplacian noise}) {$p \le 1-\Delta$} {
    $r \gets C_{\epsilon}^{-1} (p)$;
}
\Else(\tcp*[f]{uniform noise in noise circle}) {
    Choose $r$ \st $r^2$ is uniformly sampled in $[0,R^2]$\;
}
Choose $\theta \in [0,2\pi]$ uniformly at random\;
$x' \gets x + \cord{r cos \theta, r sin \theta}$\;
\KwRet{$x'$}\;
\caption{Bounded Planar Laplace (BPL)}
\label{alg:bpl}
\end{algorithm}

\algref{alg:bpl} illustrates the detailed procedure of the Bounded Planar Laplace (BPL) mechanism.
It begins by computing the radius of the noise circle $R$ based on the privacy parameters $\epsilon$ and $\delta$ in lines 1-2.
Then, a random $p$ is uniformly chosen from $[0,1]$.
If $p$ is less than a threshold $1-\Delta$, a noise is generated using $p$, following the standard Planar Laplace mechanism (\equref{eq:cdf-rev}), as shown in lines 4-5.
However, if $p$ exceeds the threshold, it implies that the size of noise generated by the standard Planar Laplace mechanism exceeds $R$.
In such cases, a uniform noise within the noise circle is selected, as demonstrated in lines 6-7.
Finally, in lines 8-10, a random angle $\theta$ is chosen, and the noise $\cord{rcos\theta, rsin\theta}$ is used to perturb $x$ and obtain $x'$.
The privacy guarantee of the BPL mechanism is proven in \lemref{lem:bpl}.

\begin{lemma}\label{lem:bpl}
The Bounded Planar Laplace (BPL) mechanism satisfies $(\epsilon,\delta)$-Geo-Indistinguishability.
\end{lemma}
\begin{IEEEproof}
(1) If $p \le 1-\Delta$, a standard planar Laplacian noise is added to the location $x$, which satisfies $\epsilon$-Geo-I \cite{andres2013geo}.

(2) If $p > 1-\Delta$, lines 6-7 fail to guarantee $\epsilon$-Geo-I. The total probability of the failure is $\Delta$, which is uniformly distributed to each location within the noise circle. Thus, the failure probability for each location is $\frac \Delta {\pi {R^2}}=\delta$.

Therefore, the BPL mechanism satisfies $(\epsilon,\delta)$-Geo-I.
\end{IEEEproof}

\begin{figure}[htbp]
\centerline{\includegraphics[width=0.5\textwidth]{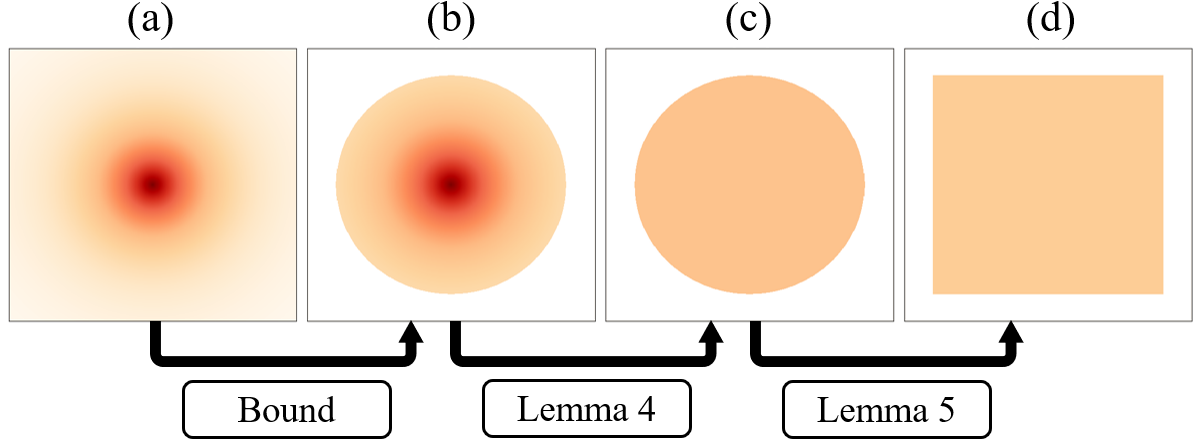}}
\caption{The probability density function (pdf) of: (a) planar Laplacian noise; (b) bounded planar Laplacian noise; (c) uniform noise in the noise circle; (d) uniform noise in the circumscribed square of the noise circle.}
\label{fig:bpl}
\end{figure}

\figref{fig:bpl}(a) and (b) show the standard planar Laplacian noise and the bounded planar Laplacian noise, respectively. The BPL noise is rigorously constrained within a size of $R$, leading to improved query performance based on our experiments.

\subsubsection{\textbf{Our Filtering Algorithm}} \label{sec:alg-filter-filter}

Our filtering performs at a grid level.
Both the query user and the data owner divide the spatial region into uniform square grids and utilize grid representations of trajectories for filtering.
Initially, the query user publishes trajectory $T_Q$ in the form of a grid sequence, denoted by $G_Q$.
Then, the data owner uses $G_Q$ to locally filter $TD$, and employs grid index to accelerate the filtering process.
The grid size should be selected carefully to ensure compliance with $(\epsilon,\delta)$-Geo-I, as discussed in \secref{sec:alg-filter-grid}.

\fakeparagraph{Query Trajectory Publishing}
Based on the Bounded Planar Laplace mechanism, we develop a novel approach for publishing query trajectory which ensures $(\epsilon,\delta)$-Geo-Indistinguishability.
The publishing algorithm involves two core operations: \textit{Perturbation} and \textit{Grid-Selection}.
Perturbation obtains a perturbed location $x'$ by adding the BPL noise to the original location, while Grid-Selection determines the grid in which $x'$ is located.

As shown in \algoref{alg:filter}, for each location $x\in T_Q$, Perturbation is executed in line 4, followed by Grid-Selection in line 5.
Subsequently, in lines 6-7, we check whether the perturbed location $x'$ and the original location $x$ are located in the same grid.
We add the grid to the candidate list $cand$ only when they are located in the same grid, ensuring that the published grids can be precisely used for filtering.
Finally, in lines 8-9, we pick $\lfloor \rho \cdot |T_Q| \rfloor$ grids from $cand$ for publishing, where $\rho$ is a ratio specified by the query user.
We prove in \thmref{thm:publish} that our publishing method satisfies the privacy constraint of $(\epsilon,\delta)$-Geo-I.

\begin{algorithm}[t]
\SetKwInOut{Input}{input}
\SetKwInOut{Output}{output}
\Input{trajectory~database~$TD$, query~trajectory~$T_Q$, privacy~parameters~$\epsilon,\delta$, publishing~rate~$\rho$}
\Output{candidate~trajectory~database~$TC$}
Choose grid size $L$ according to $\epsilon, \delta, \rho$\;
\tcp*[h]{Query user's protocol}\;
$cand \gets \phi$\;
\ForEach{location $x \in T_Q$}{
    $x' \gets$ BPL$(x, \epsilon, \delta)$\;
    $g' \gets$ \textit{grid No. of} $x'$\;
    \If{$g' =$ grid No. of $x$}
    {Add $g'$ to $cand$\;}
}
Select $\lfloor \rho \cdot |T_Q| \rfloor$ grids from $cand$ into $G_Q$, and eliminate repetitive grids in it\;
Send $G_Q$ to the data owner\;
\tcp*[h]{Data owner's protocol}\;
Construct the grid index $GI$ using $TD$\;
Upon receiving $G_Q$, compute $TC=\bigcap_{g\in G_Q}GI[g]$\;
\KwRet{$TC$}\;
\caption{$(\epsilon,\delta)$-Geo-I based Filtering}
\label{alg:filter}
\end{algorithm}

\fakeparagraph{Filtering Strategy} 
We introduce the notion of traversal grids before detailing the process of filtering the trajectory database.

\begin{definition}[Traversal Grids]
The traversal grids of a trajectory $T$ under distance threshold $\tau$, denoted as $G_{\tau}(T)$, contain all the grids covered by a set of circles $\{\textsf{circle}(x,\tau)|x\in T.locs\}$. Here $\textsf{circle}(x,\tau)$ denotes a circle centered at location $x$ and with a radius of $\tau$, and $T.locs$ represents all the locations in trajectory $T$, including intermediate locations on each segment.
\end{definition}

\begin{figure}[htbp]
\centerline{\includegraphics[width=0.48\textwidth]{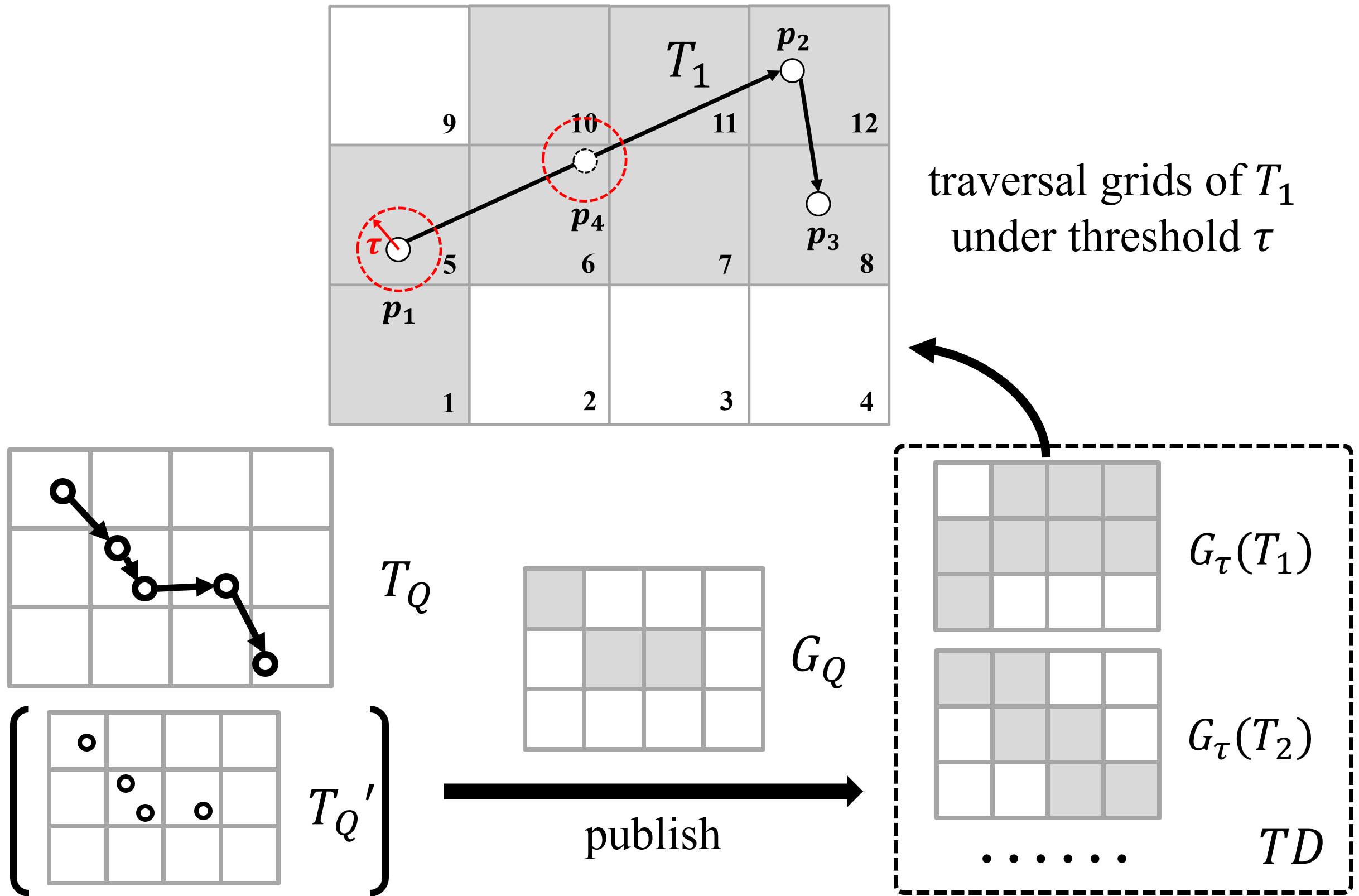}}
\caption{An illustration of Geo-I based Filtering.}
\label{fig:filter}
\end{figure}

\begin{example} \label{exp:traversal}
Consider trajectory $T_1=\{p_1,p_2,p_3\}$ in \figref{fig:filter}. Segment $s_1=(p_1,p_2)$ traverses grids $5,6,7,11,12$, and segment $s_2=(p_2,p_3)$ traverses grids $12,8$.
Besides, $\textsf{circle}(p_1,\tau)$ covers grid $1$, and $\textsf{circle}(p_4,\tau)$ ($p_4$ is a location in segment $s_1$) covers grid $10$.
Thus, the traversal grids of $T_1$ under $\tau$, $G_{\tau}(T_1)=\{1,5,6,7,8,10,11,12\}$.
\end{example}

Using the concept of the traversal grids, we formulate the filtering strategy as follow: the data owner filters $TD$ by retaining only trajectories whose traversal grids encompass all the grids in $G_Q$. The correctness of the strategy is proven in \lemref{lem:filter}.

\begin{example} \label{exp:filter}
In \figref{fig:filter}, the query user publishes $T_Q$ at a grid level for filtering. $G_Q$ is a grid sequence generated by $T_Q$ and more specifically, $T_Q$'s subtrajectory $T_Q'$.
When receiving $G_Q$, the data owner filters $TD$ according to the traversal grids of each trajectory. In our example, $T_1$ is filtered out while $T_2$ is not, since $G_Q\not\subseteq G_\tau(T_1)$ and $G_Q\subset G_\tau(T_2)$.
\end{example}

\fakeparagraph{Indexing}
Considering the substantial amount of trajectory data in the database $TD$, we introduce a grid index to accelerate our filtering strategy.
The grid index $GI$ is constructed during the offline stage.
For each trajectory in $TD$, we calculate its traversal grids and insert all the mapping relations from grid ID to trajectory ID into $GI$.
During the online stage, we can efficiently obtain the candidate database $TC$ by computing the intersection of $GI[g]$ for all $g\in G_Q$. The time for index construction is excluded from the complexity analysis since it can be finished before the query execution.

\fakeparagraph{Correctness of Our Filtering}
The correctness of the filtering strategy is proven in \lemref{lem:filter}.
\begin{lemma} \label{lem:filter}
    If the query user publishes the grid sequence $G_Q$ using trajectory $T_Q$, then $G_Q \subseteq G_{\tau}(T_i)$ is a necessary condition for $\match_{\tau}(T_i,T_Q)=\textsf{true}$.
\end{lemma}

\begin{IEEEproof}
Suppose there is a grid $g\in G_Q$ such that $g \not \in G_{\tau}(T_i)$.
Then the point $q$ in $T_Q$ that generates $g$ can never be matched by any locations in $T_i$, even when timestamps are not considered.
This implies that $\match_{\tau}(T_i,T_Q)$ should always be $\textsf{false}$ in such cases, thereby completing our proof.
\end{IEEEproof}

\fakeparagraph{Privacy of Query Trajectory Publishing} 
We prove the privacy guarantee of query trajectory publishing in \thmref{thm:publish}.
\begin{theorem} \label{thm:publish}
The query trajectory publishing algorithm (\algoref{alg:filter}) satisfies $(\epsilon,\delta)$-Geo-Indistinguishability.
\end{theorem}

\begin{IEEEproof}
We analyse the two core operations, Perturbation and Grid-Selection, respectively.
According to \lemref{lem:bpl}, Perturbation satisfies $(\epsilon,\delta)$-Geo-I.
Grid-Selection can be viewed as a post-processing after perturbation since the grid number is determined only based on $x'$ and does not rely on $x$.

\lemref{lem:post-process} has proven that post-processing does not impact the privacy guarantee. Thus, publishing location $x$ as $g'$ satisfies $(\epsilon,\delta)$-Geo-I.
Suppose $G_Q$ is generated by $T_Q'$, a subtrajectory of $T_Q$, then the procedure of publishing $T_Q'$ as $G_Q$ preserves $(\epsilon,\delta)$-Geo-I, which completes our proof.
\end{IEEEproof}

\fakeparagraph{Complexity Analysis}
Given that each entry in grid index $GI$ is sorted, the time complexity of filtering with the grid index is $O(\sum_{q\in G_Q}|GI[q]|)$, where $G_Q$ is the published grid sequence.

\vspace{0.5em}
\subsubsection{\textbf{Selection of Grid Size}} \label{sec:alg-filter-grid}
As indicated in \algoref{alg:filter}, the grid size is closely related to the privacy level and the query performance. Therefore, it is crucial to derive a proper grid size that can achieve the specified privacy level.

\begin{figure}[htbp]
\centerline{\includegraphics[width=0.3\textwidth]{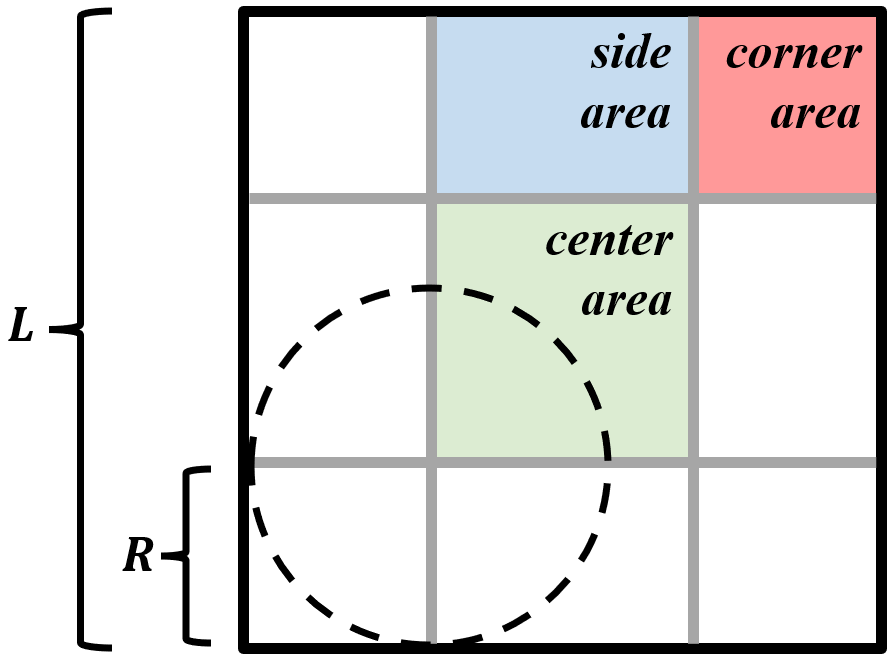}}
\caption{Division of a grid into three types of areas. A grid is a square with a side length of $L$. The radius of the noise circle is denoted as $R$ ($R \le \frac L3$).}
\label{fig:grid-area}
\end{figure}

\fakeparagraph{Basic Idea}
To determine the grid size, we need to establish a connection between the probability of successfully perturbing one location $p$, the radius of the noise circle $R$ and the grid size $L$.
Given $R$ and $L$, we can estimate $p$ by considering the probability that both location $x$ and its perturbation $x'$ fall in the same grid. 
Our analysis is based on the success probability in three different types of areas in a grid: \textit{center area}, \textit{side area}, and \textit{corner area}, as shown in \figref{fig:grid-area}. 

\fakeparagraph{Upper Bound for Grid Size $L$}
We use $p_{center}$, $p_{side}$, $p_{corner}$ to denote the probability of successful perturbation in three types of area, then the following equation holds:
\begin{equation}\label{equ:p0}
L^2p=(L-2R)^2p_{center} + 4R(L-2R)p_{side} + 4R^2p_{corner}
\end{equation}

We note that $p_{center}=1$, as for any location in the center area, the distance from the location to the grid boundary consistently exceeds $R$, the maximum size of the BPL noise.

We then proceed to analyse $p_{side}$ and $p_{corner}$. Considering the complexity of the planar Laplace distribution, we use an approximation to simplify the analysis. Specifically, we replace the BPL noise (\figref{fig:bpl}(b)) with the uniform noise in the circumscribed square of the noise circle (\figref{fig:bpl}(d)). The feasibility of this replacement will be discussed in \lemref{lem:sub1} and \lemref{lem:sub2}.

\begin{theorem} \label{thm:bound2}
Suppose $p$ represents the success probability of perturbing a single location, then a grid size $L= \frac {R} {2(1-\sqrt{p_0})}$ can ensure $p\ge p_0$.
\end{theorem}

\begin{figure}[htbp]
\centerline{\includegraphics[width=0.48\textwidth]{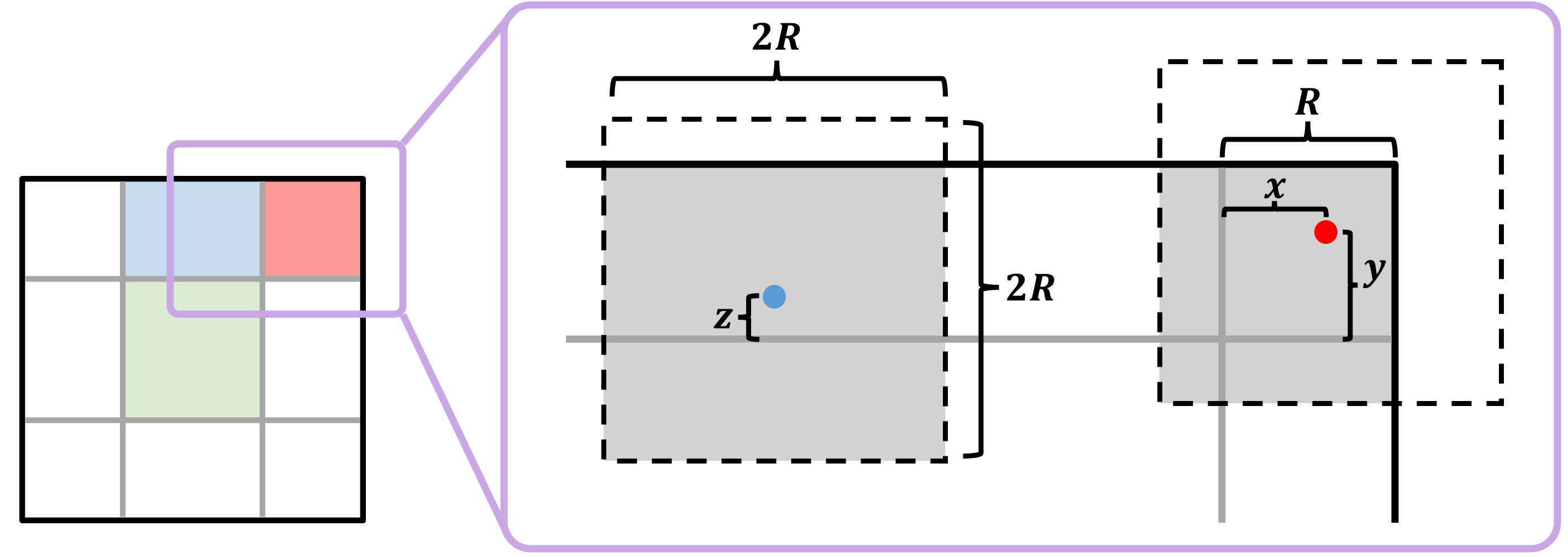}}
\caption{Using the approximation to obtain lower bounds for $p_{side}$ and $p_{corner}$. The noise is uniformly sampled in the circumscribed square of the noise circle, which is the square with a side length of $2R$. The noise in the grey area ensures the successful perturbation.}
\label{fig:approx}
\end{figure}

\begin{IEEEproof}
According to \lemref{lem:sub1} and \lemref{lem:sub2}, replacing the BPL noise with the uniform noise in the circumscribed square of the noise circle reduces the probability of successful perturbation.
Thus, we can derive the lower bounds for $p_{side}$ and $p_{corner}$:

(1) Consider the blue point in \figref{fig:approx}, $p_{side}$ satisfies:
\begin{equation*}
   p_{side} \ge \int_0^R\frac 1{4R^2}\cdot 2R(2R-z)dz=\frac 34 
\end{equation*}

(2) Consider the red point in \figref{fig:approx}, $p_{corner}$ satisfies:
\begin{equation*}
    p_{corner} \ge \int_0^{R}\int_0^{R} \frac 1{4R^2}[(2R-x)(2R-y)]\ dxdy=\frac 9{16}
\end{equation*}

Substituting $p_{center}=1,p_{side}\ge \frac 34,p_{corner}\ge \frac 9{16}$ into \equref{equ:p0}, we obtain:
\begin{align*}
L^2p &\ge (L-2R)^2+\frac 34\cdot 4R(L-2R) +\frac 9{16}\cdot 4R^2\\
\Rightarrow p &\ge (1-\frac{R}{2L})^2
\end{align*}
\vspace{0.5em}
Thus, $L=\frac {R} {2(1-\sqrt{p_0})}$ can ensure that $p \ge p_0$.
\end{IEEEproof}

\lemref{lem:sub1} and \lemref{lem:sub2} prove the feasibility of the replacement by leveraging the uniform noise in the noise circle (\figref{fig:bpl} (c)) as an intermediate.

\begin{lemma} \label{lem:sub1}
Replacing the bounded planar Laplacian noise with the uniform noise in the noise circle reduces the probability of successful perturbation.
\end{lemma}
\begin{IEEEproof}
We denote the probability of generating a BPL noise of size $x$ as $g(x)$, and the probability of generating a uniform noise of size $x$ as $u(x)$. Then we have:
\begin{equation}\label{eq:int-eq1}
\int_0^R u(x)xdx=\int_0^R g(x)xdx=1
\end{equation}

\begin{spacing}{1}
We can observe that $u(x)\equiv \frac 2{R^2}$, and $g(x)$ is monotonically decreasing in $[0,R]$.
Then we use $f(x)$ to represent the average probability of successful perturbation when the noise size is $x$.
This function is monotonically decreasing in $[0,R]$, as a larger noise reduces the success probability.
Based on the monotonicity of $f(x)$ and $g(x)$, we observe that for all $x,y\in[0,R],x\not=y$, $[f(x)-f(y)][g(x)-g(y)] > 0$, indicating that $[f(x)-f(y)][g(x)-g(y)]xy \ge 0$, hence:
\end{spacing}
\begin{small}\begin{align*}
0&\le \int_0^R\int_0^R[f(x)-f(y)][g(x)-g(y)]xydxdy\\
&=\int_0^R f(x)g(x)xdx \int_0^R ydy + \int_0^R f(y)g(y)ydy \int_0^R xdx\\
&\quad -\int_0^R\int_0^Rf(x)g(y)xydxdy-\int_0^R\int_0^Rf(y)g(x)xydxdy
\end{align*}\end{small}

According to the symmetry under double integral interchange, $\int_0^R\int_0^Rf(x)g(y)xydxdy=\int_0^R\int_0^Rf(y)g(x)xydxdy$, thus,
\begin{small}\begin{align*}
0&\le \frac {R^2} 2 \cdot \int_0^R f(x)g(x)xdx + \frac {R^2} 2 \cdot \int_0^R f(y)g(y)ydy\\
&\quad -2\cdot \int_0^R\int_0^Rf(x)g(y)xydxdy
\end{align*}\end{small}
\vspace{-1.0em}
\begin{small}\begin{align*}
\Rightarrow \frac {R^2} 2 \cdot \int_0^R f(x)g(x)xdx
&\ge \int_0^R\int_0^Rf(x)g(y)xydxdy\\
&=\int_0^Rf(x)xdx\int_0^Rg(y)ydy
\end{align*}\end{small}
since \equref{eq:int-eq1} indicates that $\int_0^R g(y)ydy=1$ and $u(x)\equiv \frac 2{R^2}$, we can obtain:
\begin{small}\begin{equation*}
\int_0^R f(x)g(x)xdx \ge \frac 2{R^2} \int_0^Rf(x)xdx=\int_0^Rf(x)u(x)xdx
\end{equation*}\end{small}

where $\int_0^R f(x)u(x)xdx$ is the success probability when using the uniform noise in the noise circle, and $\int_0^R f(x)g(x)xdx$ is the success probability when using the BPL noise, thus \lemref{lem:sub1} holds true.
\end{IEEEproof}

\begin{lemma} \label{lem:sub2}
Replacing the uniform noise in the noise circle with the uniform noise in the noise circle's circumscribed square reduces the probability of successful perturbation.
\end{lemma}

\begin{IEEEproof}
We prove \lemref{lem:sub2} by deriving the following inequation in \figref{fig:lem}:
\begin{equation}\label{eq:lem}
\frac {C_{\cap}}C \ge \frac {S_{\cap}}S
\end{equation}

\begin{figure}
    \centering
    \subfloat[Inside Circle]{
        \includegraphics[width=0.16\textwidth]{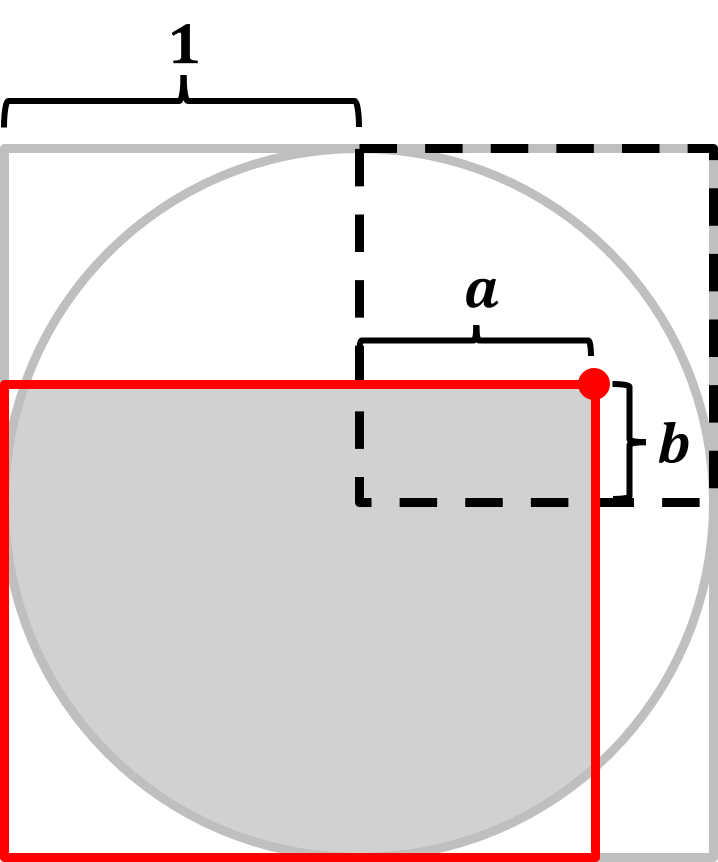}
    }
    \hspace{5mm}
    \subfloat[Outside Circle]{
        \includegraphics[width=0.16\textwidth]{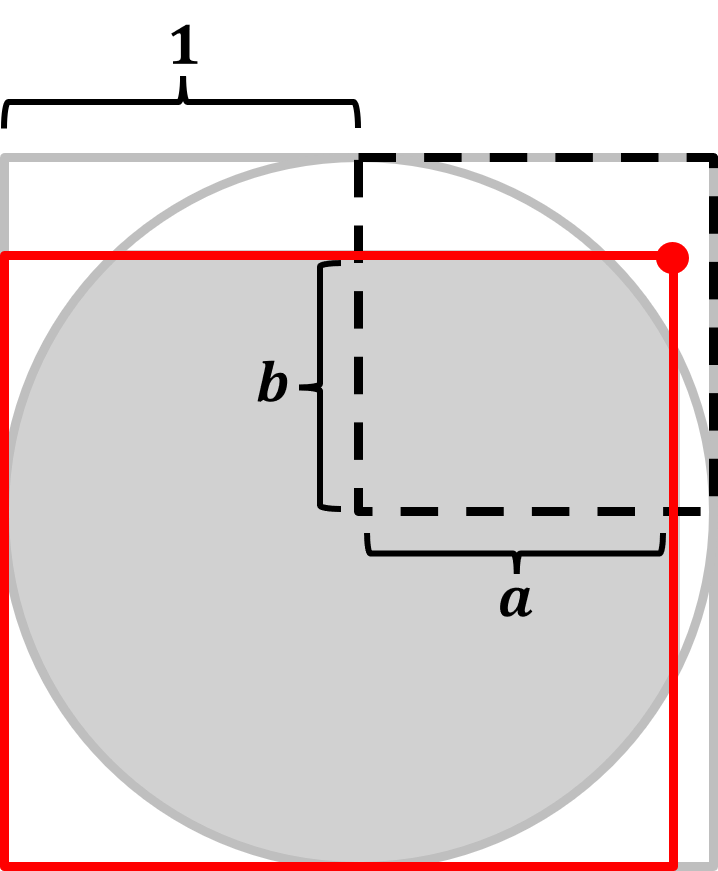}
    }
    \caption{WLOG, we assume the radius of the noise circle to be $1$. The upper right point of the red rectangle is located within the dashed box. $C_{\cap}$ denotes the size of the grey area (intersection between the noise circle and the red rectangle), and $C$ denotes the size of the noise circle. $S_{\cap}$ denotes the size of the red rectangle, and $S$ denotes the size of the large square.}
    \label{fig:lem}
\end{figure}

(1) If the random location lies inside the noise circle ($\sqrt{a^2+b^2}\le1$), as shown in \figref{fig:lem}(a):
\begin{small}\begin{equation*}
C_{\cap}=\frac \pi 4+ab+\frac 12(a\sqrt {1-a^2}+b \sqrt{1-b^2} +arcsin\ a+arcsin\ b)
\end{equation*}\end{small}

(2) If the random location lies outside the noise circle ($\sqrt{a^2+b^2}>1$), as shown in \figref{fig:lem}(b):
\begin{small}\begin{equation*}
C_{\cap}=a\sqrt{1-a^2}+b\sqrt{1-b^2}+arcsin\ a+arcsin\ b
\end{equation*}\end{small}

Besides, we have $S_{\cap}=(1+a)(1+b)$, $C=\pi$, $S=4$. It can be confirmed that for all $a,b\in[0,1]$, \equref{eq:lem} holds true, which completes our proof.
\end{IEEEproof}

\subsection{SMC based Verification} \label{sec:alg-verify}
After filtering, we reduce the search space to a smaller candidate database $TC$, where each trajectory traverses all grids in the published grid sequence $G_Q$.
However, as SMC operations are typically slower than their plaintext counterparts \cite{lindell2020secure}, performing SMC over the potentially large $TC$ is time-consuming.
Thus, we introduce a data partition scheme along with a reference trajectory based pruning strategy to further improve efficiency.

\begin{figure}[htbp]
\centerline{\includegraphics[width=0.48\textwidth]{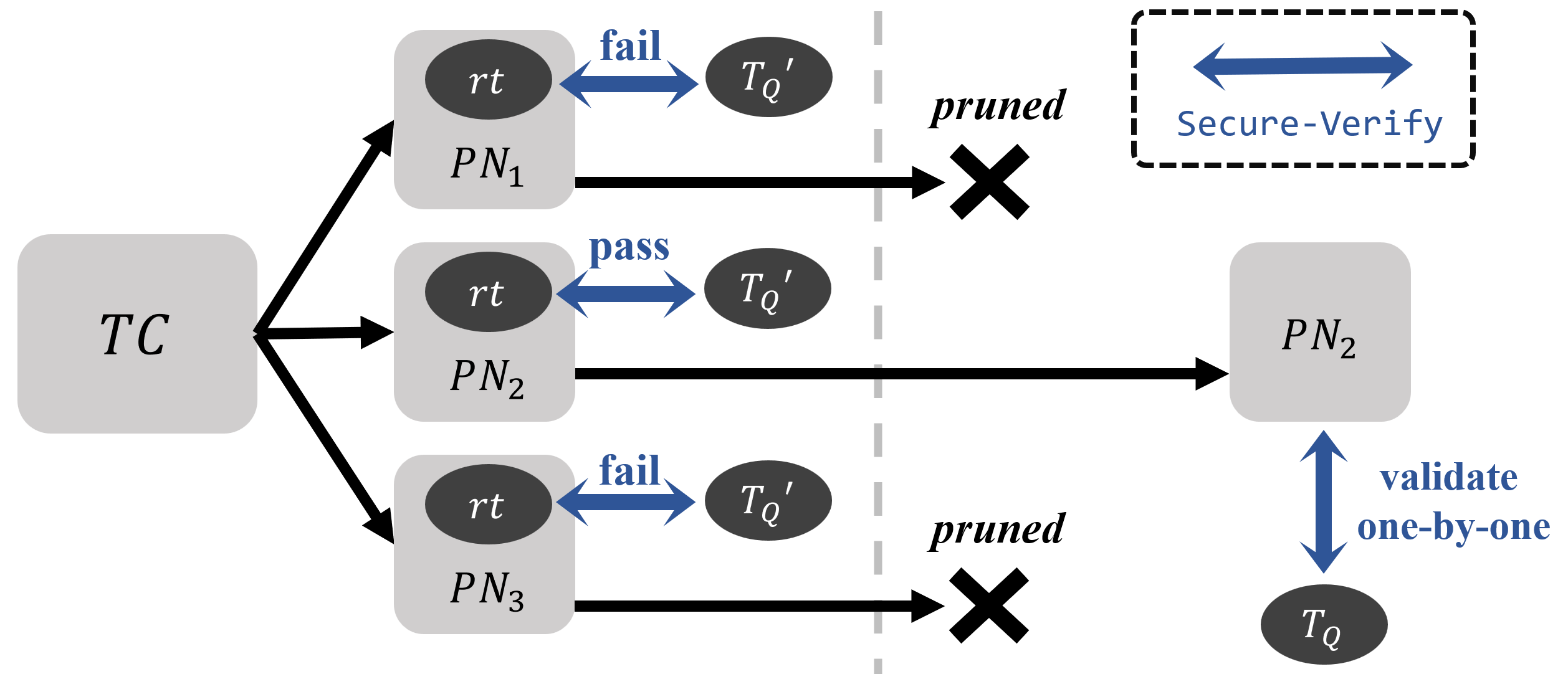}}
\vskip -4pt
\caption{An illustration of SMC based Verification.}
\label{fig:verify}
\end{figure}

\fakeparagraph{Basic Idea} 
The idea of SMC based verification is illustrated in \figref{fig:verify}.
We devide the candidate database $TC$ into multiple data partitions $PN_i$ based on the spatiotemporal characteristics of trajectories.
For each partition $PN_i$, we generate a special trajectory, termed the \textit{reference trajectory} $rt$, which encapsulates the spatiotemporal features of all trajectories in $PN_i$.
We then apply \lemref{lem:pruning} to prune partitions where none of the trajectories can match $T_Q$, thus avoiding the need to validate each trajectory in $TC$ through SMCs.

\begin{algorithm}[t]
\SetKwInOut{Input}{input}
\SetKwInOut{Output}{output}
\Input{candidate~trajectory~database~$TC$, query~trajectory~$T_Q$, distance~threshold~$\tau$, grid~size~$L$}
\Output{all trajectories that matches $T_Q$}
\SetKwFunction{SV}{Secure-Verify}
$result \gets \phi$\;
Partition $TC$ with $m \gets \lfloor \alpha \sqrt{|TC|} \rfloor$ to obtain $PNs$\;
\ForEach{partition $PN \in PNs$}{
    Generate $rt$ as the reference trajectory of $PN$\;
    \tcp{Pruning}
    $match_{rt} \gets$ \SV{$rt, T_Q', \tau+\sqrt{2} L$}\;
    \If{$match_{rt}$ is $\textsf{false}$}{
        \continue
    }
    \tcp{Final Validation}
    \ForEach{trajectory $t\in PN$}{
        $match_t \gets$ \SV{$t, T_Q, \tau$}\;
        \If {$match_t$ is $\textsf{true}$}{
            Add $t$ to $result$\;
        }
    }
}
\KwRet{$result$}\;
\vspace{0.6em}

\SetKwProg{SF}{Function}{:}{\KwRet}
\SF{\SV{$T$, $T_Q$, $\tau$}}{
    $N \gets 0$\;
    \ForEach{point $q \in T_Q$}{
        $match_q \gets 0$\;
        \ForEach{segment $s \in T$}{
            Compute $loc_s(ts)$, which derives the location of $s$ at timestamp $ts$\;
            \underline{$dis \gets \dist(q.loc,\ loc_s(q.ts))$}\;
            \If{\underline{$dis \le \tau$ and $q.ts \in [s.o.ts, s.d.ts]$}}{
                \underline{$match_q \gets 1$}\;
            }
        }
        \underline{$N \gets N + match_q$}\;
    }
    \underline{$match \gets N$ = $|T_Q|$}\;
    \color{black}
    \KwRet{$match$}\;
}
\caption{SMC based Verification}
\label{alg:verify}
\end{algorithm}

\fakeparagraph{Data Partition} 
Based on the spatiotemporal information of trajectories, we divide the database $TC$ into multiple partitions, each containing no more than $m$ trajectories. We consider trajectories that traverse a grid at approximately the same time as similar and group them into the same partition. 
Specifically, the data partition is performed as follows. We begin by grouping all trajectories in $TC$ into a single partition.
Subsequently, this partition undergoes subdivision through the splitting operation until each resulting partition has a size smaller than $m$.

The splitting operation for $PN$ is based on the timespan of the trajectories within it.
We select the grid with the longest timespan as the splitting criteria, denoted as $g_{split}$. This means trajectories in $PN$ are separated into different partitions based on when they traverse grid $g_{split}$.
Then the splitting value, $v_{split}$, is set to be the median of the ending times for all trajectories in partition $PN$. As a result, $PN$ can be divided into two partitions of equal sizes or sizes differing by $1$.

\begin{figure}[htbp]
\centerline{\includegraphics[width=0.48\textwidth]{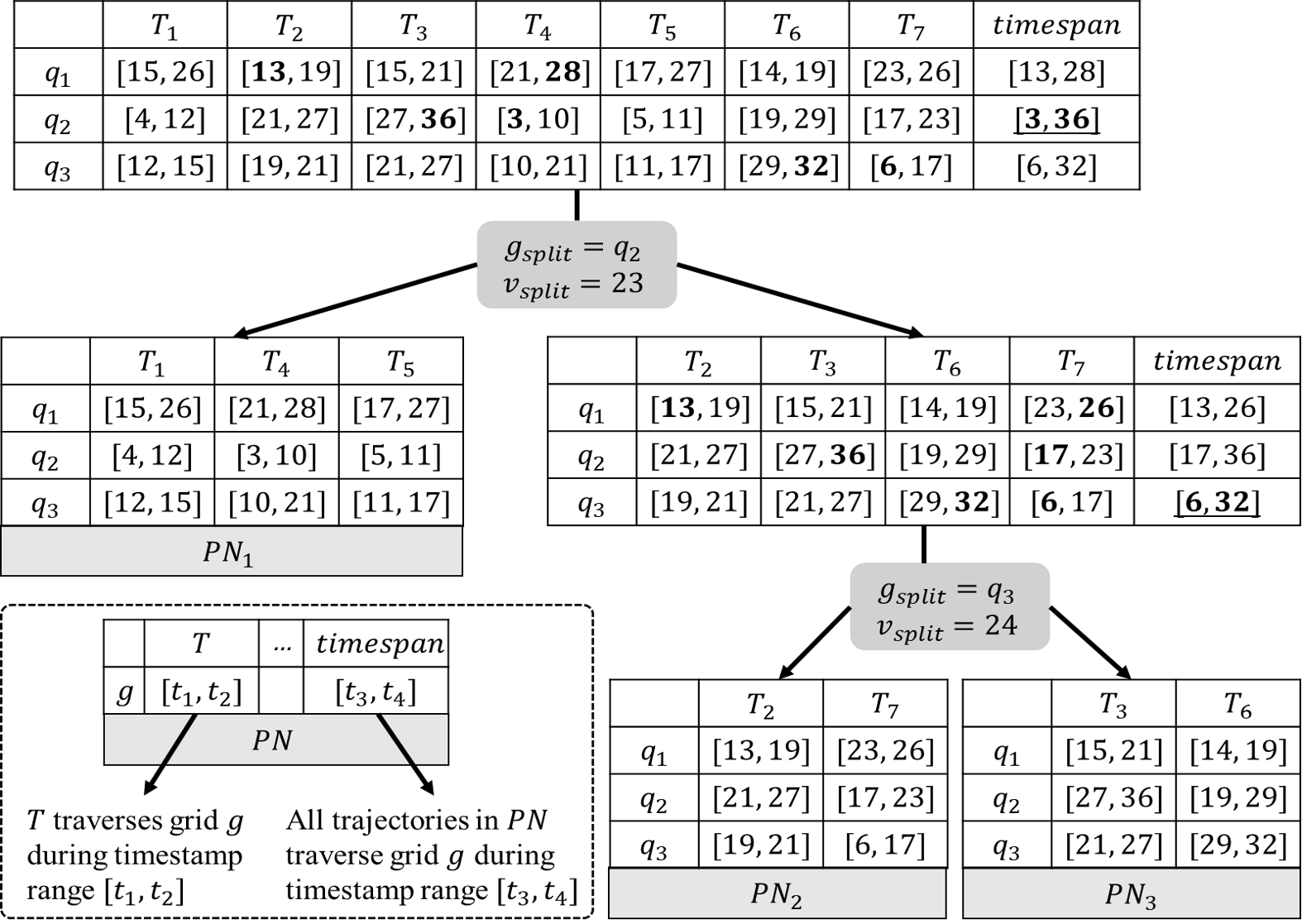}}
\caption{Divide database $TC=\{T_1,...,T_7\}$ into $3$ partitions. All the trajectories in $TC$ traverse grids $q_1$, $q_2$ and $q_3$.}
\label{fig:partition}
\end{figure}

\fakeparagraph{Pruning with Reference Trajectory} 
We use the reference trajectory within each partition for pruning. The reference trajectory is generated by exploring all trajectory points within grid $g$ for each $g \in G_Q$. The point with the smallest timestamp is selected as the original point $o$, and the one with the largest timestamp is selected as the destination point $d$. The segment formed by $o$ and $d$ is added to the reference trajectory $rt$. This process is repeated for each grid $g \in G_Q$, resulting in a reference trajectory with $|G_Q|$ segments. Then $rt$ can be used for pruning according to \lemref{lem:pruning}.

\fakeparagraph{Algorithm Details}
The procedure of SMC based verification is presented in \algoref{alg:verify}.
In line 2, data partition is performed on the candidate trajectories $TC$, with the partition size limited to $\lfloor \alpha \sqrt{|TC|} \rfloor$. The parameter $\alpha$ is adjustable and discussed in Section \ref{sec:exp-ablation}.
Subsequently, in line 4, a reference trajectory $rt$ is generated for each partition, and lines 5-7 utilize it for pruning based on Lemma \ref{lem:pruning}.
If $rt$ does not match $TQ'$ under the threshold of $\sqrt 2 L + \tau$, all trajectories in its partition are pruned. Following pruning, every trajectory in the remaining partitions undergoes final validation via \texttt{Secure-Verify} to ascertain whether they match $T_Q$, as shown in lines 8-11.

The \texttt{Secure-Verify} function in \algoref{alg:verify} utilizes SMC techniques to verify whether a trajectory $T$ matches $T_Q$ based on \defref{def:match}.

In lines 15-17, each point $q \in T_Q$ is iterated through to check if it can be matched by at least one segment in $T$.
For each segment $s$, the data owner locally derives a function $loc_s(ts)$ for estimating the location of $s$ at timestamp $ts$ (line 18).
In line 19, the Euclidean distance $dis$ between $q.loc$ and its corresponding point in $s$ is securely computed.
If $dis$ is less than $\tau$ and the sequential order of timestamps is satisfied, $match_q$ is set to $1$, indicating that $q$ can be matched by $s$ (lines 20-21).
If at least one segment matches $q$, the total matching number $N$ increases by $1$ (line 22).
Lines 23-24 securely compare the total matching number with the length of $T_Q$ and output the result indicating whether $T$ matches $T_Q$ or not.
The underlined code segments need to be implemented with two-party SMC protocols (\eg by using Obliv-C \cite{oblivc}).

\fakeparagraph{Correctness of Our Pruning}
The correctness of pruning in lines 5-7 of \algoref{alg:verify} is proven in \lemref{lem:pruning}.
\begin{lemma} \label{lem:pruning}
If the reference trajectory $rt$ of partition $PN$ fails to match $T_Q'$ under a threshold of $\tau + \sqrt{2} L$ ($L > \tau$), then no trajectory in $PN$ can match $T_Q$ under a threshold of $\tau$. Here, $T_Q'$ denotes the subtrajectory of $T_Q$ that generates $G_Q$, and $L$ is the side length of a grid.
\end{lemma}

\begin{IEEEproof}
We prove \lemref{lem:pruning} by demonstrating that if a point $q$ in $T_Q'$ cannot be matched by any segments in $rt$ under a threshold of $\sqrt{2} L + \tau$, it cannot either be matched by any trajectories in $PN$ under a threshold of $\tau$.

Consider the case where the distance from $q$ to the grid boundary is always larger than $\tau$, implying that $q$ can only be matched by locations within the same grid. Since $\sqrt{2} L$ is the longest distance between two locations within a grid, none of the segments in $rt$ traverse $q$'s grid at timestamp $q.ts$. Consequently, no trajectory in $PN$ traverses $q$'s grid at timestamp $q.ts$, indicating that $q$ cannot be matched by any trajectory in $PN$. 
We can extend this conclusion to the general case by raising the threshold from $\sqrt{2} L$ to $\sqrt{2} L + \tau$, since $L > \tau$ indicates that $q$ can only be matched by the point from the adjacent grid, thus completing our proof.
\end{IEEEproof}

\fakeparagraph{Security of SMC Based Verification}
In the SMC based verification, the number and length of reference trajectories and trajectories in the remaining partitions are disclosed to facilitate the execution of \texttt{Secure-Verify}.
Apart from this, all other information regarding trajectories in $TC$ and query trajectory $T_Q$ are thoroughly protected by SMC in the semi-honest model.

\fakeparagraph{Complexity Analysis}
The complexity of pruning and final validation is $O( |T_Q'|\cdot|G_Q|\cdot\frac {|TC|}{m}+|T_Q|\cdot|T|\cdot n_r m)$, where $n_r$ is the number of partitions surviving the pruning.
According to the experiments on real-world datasets, $n_r$ can be regarded as a constant. Besides, $|T_Q|$, $|T_Q'|$, $|G_Q|$ and $|T|$ are constants related to the trajectory length. Therefore, we choose $m=\Theta(\sqrt{|TC|})$ (\ie $m=\lfloor \alpha \sqrt{|TC|} \rfloor$) in line 2 of \algoref{alg:verify} to achieve optimal complexity.
\section{Experimental Study} \label{sec:exp}

This section presents our experimental evaluation.
We first introduce the experiment setup (\secref{sec:exp-setup}). Then, we present the performance on real datasets (\secref{sec:exp-real}), scalability tests (\secref{sec:exp-scale}), ablation studies (\secref{sec:exp-ablation}), and extension to multiple data owners (\secref{sec:exp-fed}). Finally, we summarize the major experimental findings (\secref{sec:exp-summary}).

\subsection{Experiment Setup}\label{sec:exp-setup}
\fakeparagraph{Datasets}
We use five real-world trajectory datasets.
\begin{itemize}
    \item \textbf{Geolife \cite{zheng2010geolife}.} 
    It contains daily trajectories of individuals collected by MSRA from April 2007 to August 2012.
    \item \textbf{Dazhong \cite{liu2019social}.} 
    It contains trajectories of 13,013 cars collected by SAIC Volkswagen \cite{vw} in April 2016 and May 2016.
    \item \textbf{Xi'an \& Chengdu \cite{didi}.} 
    They are trajectory datasets published by Didi Chuxing's ride-hailing services in Xi'an and Chengdu, respectively, during October 2016.
    \item \textbf{Multi-Company \cite{tong2022hu}.} 
    It is a shared trajectory dataset from 5 taxi companies in Beijing (\eg JinYinJian \cite{jinyijian}). Each company can be naturally regarded as a data owner of its collected trajectories.
\end{itemize}

\begin{table}[t]\footnotesize
    \centering
    \caption{Real datasets.}\label{tab:dataset}
    \vspace{-1ex}
    \begin{tabular}{cccccc}
    \toprule
    \textbf{Dataset} & Geolife & Dazhong & Xi'an & Chengdu & Multi-Company  \\ \midrule
    \textbf{$|TD|$}  & 11k     & 700k    & 3200k & 6000k   & 2400k  \\
    \textbf{Size}    & 0.3G    & 1.6G    & 10.9G & 16.7G   & 38.7G  \\
    \bottomrule
    \end{tabular}
\end{table}

\begin{table}[t]\footnotesize
    \centering
    \caption{Parameter settings.}\label{tab:para}
    \vspace{-1ex}
    \begin{tabular}{cc}
    \toprule
    \textbf{Parameter} & \textbf{Setting}  \\ \midrule
    Sampling Rate     & 5\%, 10\%, 20\%, 40\%  \\
    Trajectory Scalability $|TD|$    & 1500k, 3000k, 4500k, 6000k  \\
    Privacy Budget $\epsilon$  & 0.01, 0.02, 0.03, 0.04, 0.05 \\
    Partition Parameter $\alpha$     & 0.25, 0.5, 1, 2, 4  \\
    \#(Data Owner)     & 1, 2, 3, 4, 5  \\
    \bottomrule
    \end{tabular}
\end{table}

\fakeparagraph{Baselines}
We compare our framework \frameworkabbr with the following SMC based solutions.
\begin{itemize}
    \item \textbf{OblivC \cite{oblivc}.} It uses secure multi-party computation to determine whether two trajectories match or not.
    \item \textbf{STSC-ext \cite{liu2015efficient}.} It uses both addictively homomorphic encryption and secure multi-party computation to calculate the similarity between trajectories. When extending this method to \problemabbr, we assume that timestamps of points in $T_Q$ are published first. In contrast, the timestamp information does not have to be published in \frameworkabbr.
\end{itemize}

Moreover, we compare the filtering effectiveness of \frameworkabbr with existing works on trajectory publishing \cite{andres2013geo,cunningham2021real,zhang2023trajectory} in \secref{sec:exp-ablation}.
Among them, GeoI \cite{andres2013geo} is a seminal location protection mechanism that has been widely used in existing works. NGram \cite{cunningham2021real} and ATP \cite{zhang2023trajectory} are state-of-the-art solutions for privately trajectory publication.

\fakeparagraph{Metrics}
We mainly assess the efficiency of \frameworkabbr and baselines by the following metrics.
\begin{itemize}
    \item \textbf{Running time.} It is the average response time for answering one \problemabbr query.
    \item \textbf{Communication cost.} It is the total network communication between the query user and all data owners.
    \item \textbf{Retention rate.} It is the ratio of the size of the candidate trajectory database $TC$ to the original database $TD$. The retention rate ranges between $0$ and $1$, with a smaller rate indicating a more effective filtering method.
\end{itemize}
Apart from the above three metrics, we also report the index size of our framework \frameworkabbr in \secref{sec:exp-scale}.

\fakeparagraph{Implementation}
All the methods are implemented in C/C++ and compiled using GCC/G++ 9.4.0. We employ Obliv-C \cite{oblivc} for SMC operations and GMP library \cite{gnump} for big integer computation.
For all methods, we set the distance threshold $\tau=50m$.
In \frameworkabbr, we set publishing rate $\rho=60\%$, privacy parameter $\delta=10^{-5}$, and partition parameter $\alpha=0.5$. 
The query trajectory is randomly sampled from the dataset $TD$, and we reserve a subset of points in it.
The portion of reserved points is controlled by the \textit{sampling rate} parameter in \tabref{tab:para}.
We also vary other parameters, including trajectory scalability $|TD|$, privacy budget $\epsilon$, partition parameter $\alpha$, and the number of data owners.
For each parameter setting, the average result of 100 queries is reported.

\fakeparagraph{Environment}
Experiments are carries out on 5 servers connected by LAN, each with 24 2.60GHz Intel(R) Xeon(R) Platinum 8361HC CPU processors and 128GB of memory.

\subsection{Experiments on Real Datasets}\label{sec:exp-real}

To illustrate the efficiency of \frameworkabbr in real-world applications, we conduct experiments on three real datasets, Geolife \cite{zheng2010geolife}, Dazhong \cite{vw} and Xi'an \cite{didi}.

\begin{figure}
    \centering
    \begin{subfigure}{0.48\textwidth}
        \centering
        \includegraphics[width=\textwidth]{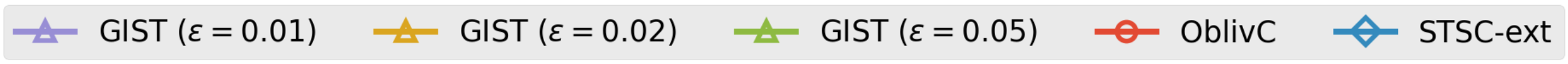}
        \caption*{}
    \end{subfigure}
    \vskip -16pt
    \begin{subfigure}{0.48\textwidth}
        \centering
        \includegraphics[width=\textwidth]{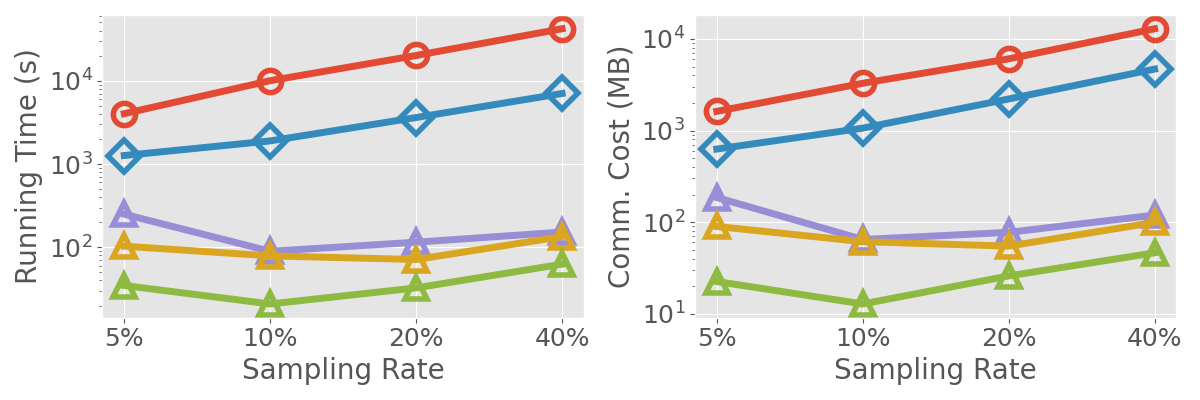}
        \vskip -6pt
        \caption{Running time and communication cost on Geolife}
    \end{subfigure}
    \begin{subfigure}{0.48\textwidth}
        \centering
        \includegraphics[width=\textwidth]{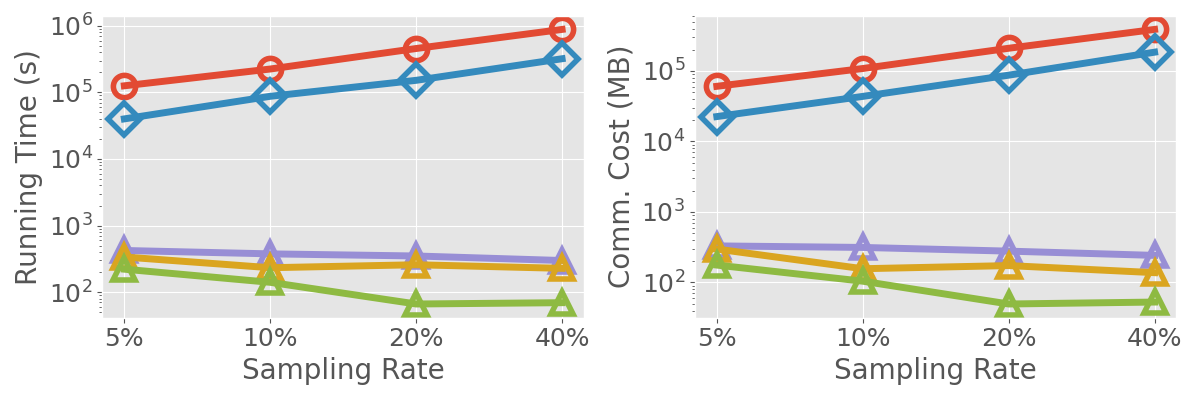}
        \vskip -6pt
        \caption{Running time and communication cost on Dazhong}
    \end{subfigure}
    \begin{subfigure}{0.48\textwidth}
        \centering
        \includegraphics[width=\textwidth]{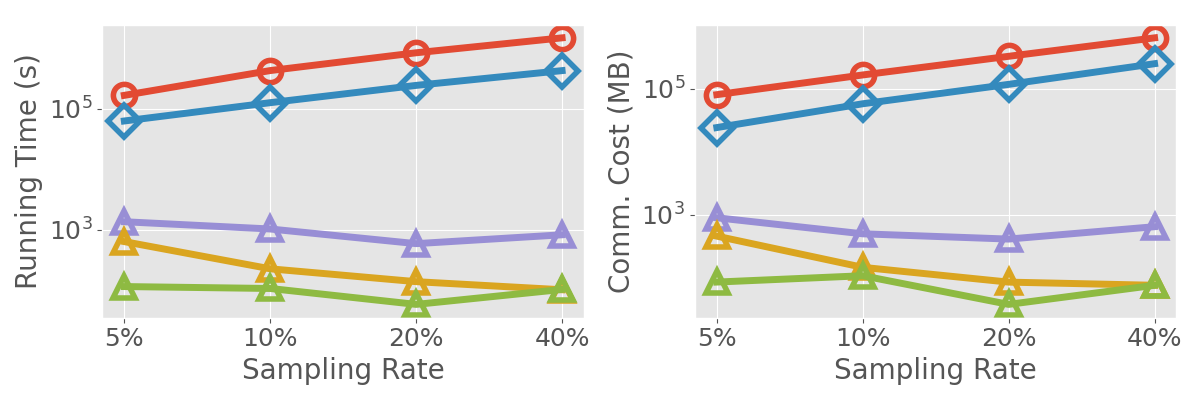}
        \vskip -6pt
        \caption{Running time and communication cost on Xi'an}
    \end{subfigure}
    \caption{Running time and communication cost of \problemabbr under different sampling rates of $T_Q$.}
    \label{fig:exp-real}
\end{figure}

\fakeparagraph{Comparison Across Datasets}
\figref{fig:exp-real} shows the running time and communication cost of \problemabbr in these real datasets.
Four sampling rate are used: $5\%$, $10\%$, $20\%$ and $40\%$. Across all three real datasets, \frameworkabbr consistently outperforms the OblivC and the STSC-ext.
In the Xi'an dataset \cite{didi} which contains 3.2 million trajectories, \frameworkabbr ($\epsilon=0.01$) is 46.6× to 418.4× faster than the runner-up STSC-ext, while incurring 2 to 3 magnitudes lower communication cost.
The privacy budget $\epsilon$ notably impact the performance of \frameworkabbr. A relatively weaker privacy preservation level, represented by a larger $\epsilon$, allows a finer granularity of the grid, leading to improved filtering effectiveness and reduced cost.

\fakeparagraph{Vary Sampling Rate}
The efficiency of \frameworkabbr is primarily influenced by two factors: the number of trajectories requiring verification and the cost associated with securely verifying each trajectory.
As the sampling rate increases, the effectiveness of filtering improves, leading to a reduction in the number of trajectories that need verification.
Meanwhile, the cost of verifying each trajectory also increases. We observe that in Geolife dataset, the efficiency of \frameworkabbr ($\epsilon=0.01$) peaks when the sampling rate is $10\%$, while in Dazhong and Xi'an datasets, \frameworkabbr ($\epsilon=0.01$) achieves the best performance when the sampling rate is $40\%$ and $20\%$, respectively.

\begin{figure*}
    \centering
    \begin{subfigure}{0.48\textwidth}
        \centering
        \includegraphics[width=\textwidth]{figs/exp/legend1.png}
        \caption*{}
    \end{subfigure}
    \vskip -16pt
    \begin{subfigure}{0.48\textwidth}
        \centering
        \includegraphics[width=\textwidth]{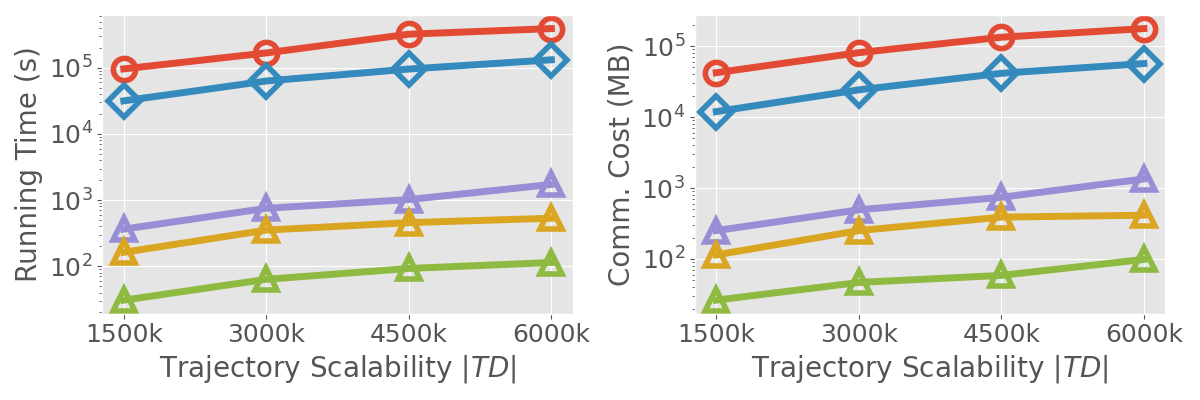}
        \vskip -6pt
        \caption{Sampling rate of 5\%}
    \end{subfigure}
    \begin{subfigure}{0.48\textwidth}
        \centering
        \includegraphics[width=\textwidth]{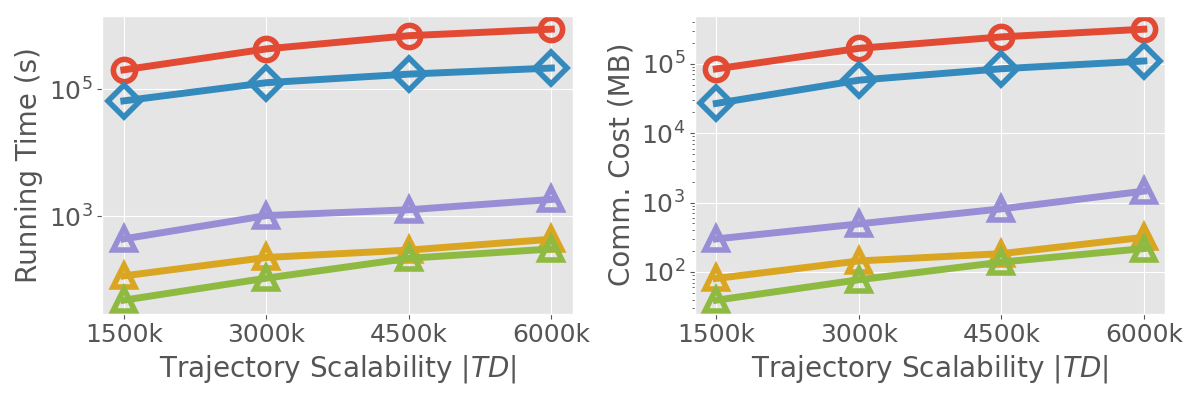}
        \vskip -6pt
        \caption{Sampling rate of 10\%}
    \end{subfigure}

    \begin{subfigure}{0.48\textwidth}
        \centering
        \includegraphics[width=\textwidth]{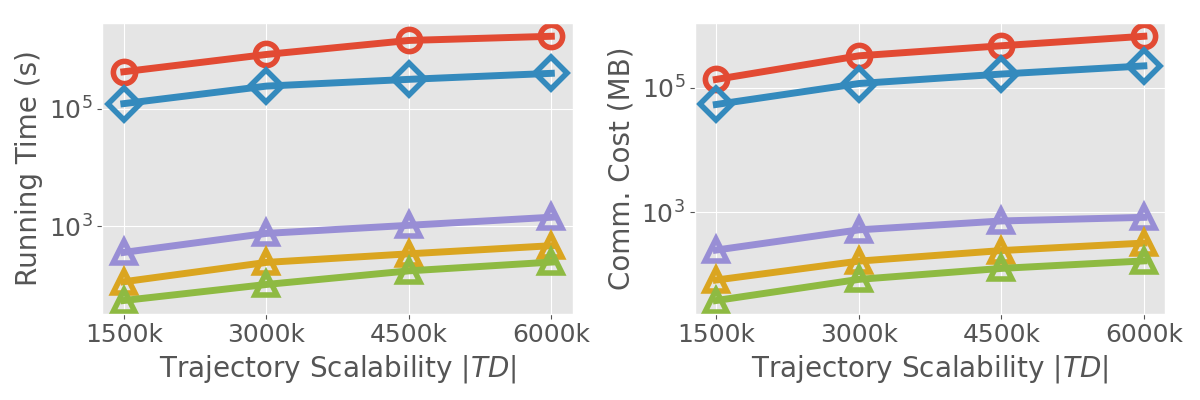}
        \vskip -6pt
        \caption{Sampling rate of 20\%}
    \end{subfigure}
    \begin{subfigure}{0.48\textwidth}
        \centering
        \includegraphics[width=\textwidth]{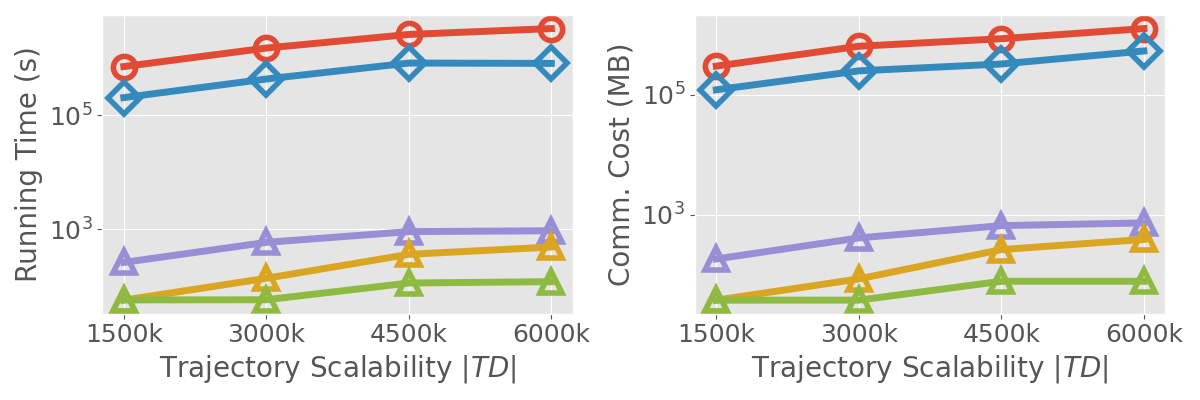}
        \vskip -6pt
        \caption{Sampling rate of 40\%}
    \end{subfigure}
    \caption{Running time and communication cost of scalability tests on the Chengdu dataset.}
    \label{fig:exp-scale}
\end{figure*}

\subsection{Experiments on Scalability Test}\label{sec:exp-scale}
We use Chengdu dataset \cite{didi} for scalability tests.
Trajectory datasets $TD$ of different sizes are generated by randomly sampling from the original dataset.

\fakeparagraph{Vary Trajectory Scalability $|TD|$}
\figref{fig:exp-scale} presents the results of scalability test under four levels of trajectory scalability $|TD|$. Generally, the running time and communication cost of \frameworkabbr grow linearly with the data size, and maintain an advantage of 2 to 3 orders of magnitude over the best baseline at all sampling rates. Using sampling rate of 20\% as an example, \frameworkabbr ($\epsilon=0.01$) is 286.5× to 348.8× more efficient than STSC-ext, and takes 222.1× to 276.4× lower communication.

\fakeparagraph{Vary Sampling Rate}
As the sampling rate increases, the advantages of \frameworkabbr become more evident. Taking trajectory scalability of 6 million as an example, \frameworkabbr ($\epsilon=0.01$) is 42.5× more efficient than the baseline when the sampling rate is 5\%, and 526.8× more efficient when the sampling rate is 40\%.
This is because the baseline methods suffer from a prominent performance loss as the length of $T_Q$ increases.
In contrast, a longer query trajectory can be leveraged in \frameworkabbr to enhance the filtering effectiveness, compensating for the performance loss in verification.

\fakeparagraph{Index Size}
\tabref{tab:index} presents the size of the grid index in \frameworkabbr under different trajectory scalabilities.
We observe that the grid index's size (585.2MB) is far more smaller than the size of the raw trajectory data (16.7G).
Overall, the space cost of the index is acceptable, considering the memory size of a modern server.

\begin{table}[t]\footnotesize
    \centering
    \caption{Size of grid index in \frameworkabbr ($\epsilon=0.01$) under different trajectory scalabilities.}\label{tab:index}
    \vspace{-1ex}
    \begin{tabular}{ccccc}
    \toprule
    \textbf{Trajectory Scalability $|TD|$} & 1500k & 3000k & 4500k & 6000k  \\ \midrule
    \textbf{Index Size (MB)}               & 154.8 & 321.7 & 448.1 & 585.2   \\
    \bottomrule
    \end{tabular}
\end{table}

\subsection{Experiments on Ablation Study}\label{sec:exp-ablation}

We conduct ablation studies from two aspects of our \frameworkabbr: the \textit{filtering} and \textit{verification} phases. The following experiments are performed on Dazhong dataset \cite{vw}, using $10\%$ and $40\%$ as the sampling rate for $T_Q$.

\subsubsection{\textbf{Ablation Study on Privacy Mechanism in Filtering}} \label{sec:exp-ablation-filter}
We compare our privacy mechanism BFL with baselines for location or trajectory privacy protection in \frameworkabbr's filtering phase, including GeoI \cite{andres2013geo}, NGram \cite{cunningham2021real} and ATP \cite{zhang2023trajectory}.
In the context of the filtering phase, we apply these baselines by publishing the query trajectory $T_Q$ as $T_Q'$ and using a safe threshold $\mathcal{T}$ for filtering.
The safe threshold ensures that if the distance between trajectory $T$ and $TQ'$ is larger than $\mathcal{T}$, then $T$ can be ruled out safely.
Specifically, $\mathcal{T}$ is calculated as the sum of the distance threshold $\tau$ and the largest distance between a point in $T_Q$ and its corresponding point in $T_Q'$.
To ensure a fair comparison with these methods, the privacy budgets $\epsilon$ in NGram and ATP are normalized by the length of $T_Q$ to maintain consistent interpretations with the other methods.

\begin{figure}
    \centering
    \begin{subfigure}{0.3\textwidth}
        \centering
        \includegraphics[width=\textwidth]{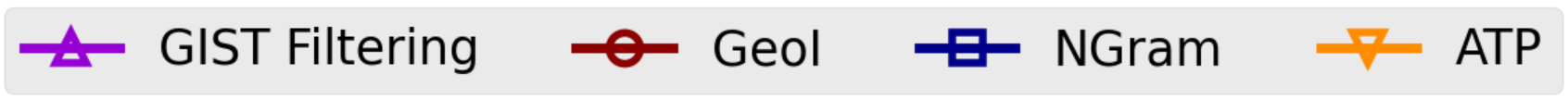}
        \caption*{}
    \end{subfigure}
    \vskip -16pt
    \begin{subfigure}{0.24\textwidth}
        \centering
        \includegraphics[width=\textwidth]{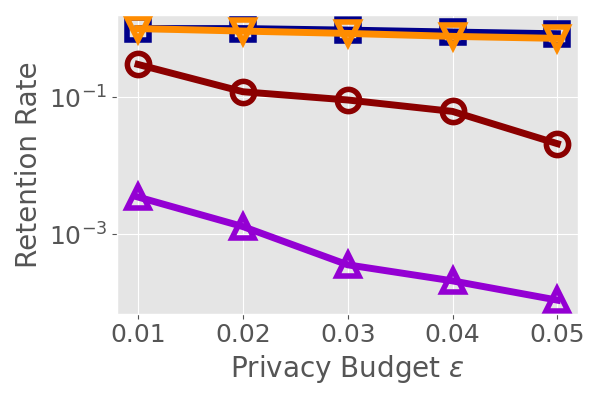}
        \vskip -6pt
        \caption{Sampling rate of 10\%}
    \end{subfigure}
    \begin{subfigure}{0.24\textwidth}
        \centering
        \includegraphics[width=\textwidth]{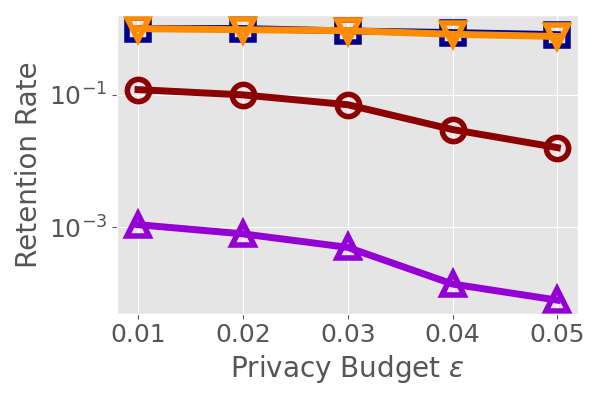}
        \vskip -6pt
        \caption{Sampling rate of 40\%}
    \end{subfigure}
    \caption{Retention rate of \frameworkabbr filtering and other trajectory publishing methods under different privacy budgets $\epsilon$.}
    \label{fig:exp-filter}
\end{figure}

\fakeparagraph{Vary Privacy Budget $\epsilon$}
\figref{fig:exp-filter} shows the filtering effectiveness of different filtering methods across varying privacy budgets.
It is demonstrated that for each method, a weaker privacy level indicated by larger $\epsilon$, proves to be more effective in preserving the utility of the query trajectory, resulting in a lower filtering rate.
Notably, our filtering method demonstrates at least an 85× improvement in effectiveness under the same privacy budget.
The results also reveal that the trajectories published by NGram and ATP fail to maintain high effectiveness when used for filtering.
These results illustrate the necessity of designing a novel privacy mechanism within our framework.

\subsubsection{\textbf{Ablation Study on Pruning in Verification}} \label{sec:exp-ablation-verify}

The performance of SMC based verification is impacted by the partition parameter $\alpha$.
We compare the verification efficiency under different choices of $\alpha$.

\begin{figure}
    \centering
    \begin{subfigure}{0.18\textwidth}
        \centering
        \includegraphics[width=\textwidth]{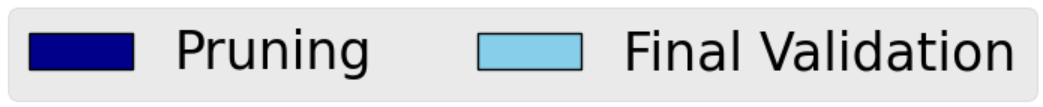}
        \caption*{}
    \end{subfigure}
    \vskip -16pt
    \begin{subfigure}{0.24\textwidth}
        \centering
        \includegraphics[width=\textwidth]{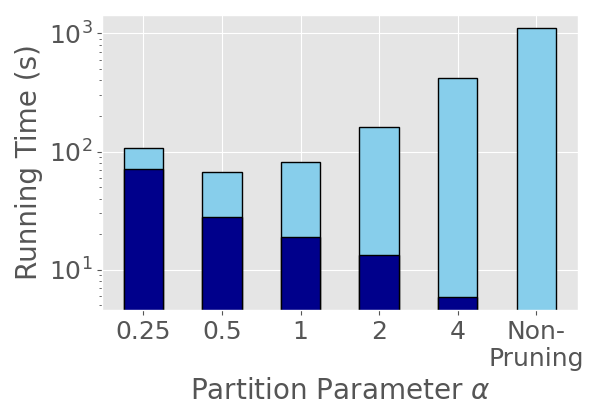}
        \vskip -6pt
        \caption{Sampling rate of 10\%}
    \end{subfigure}
    \begin{subfigure}{0.24\textwidth}
        \centering
        \includegraphics[width=\textwidth]{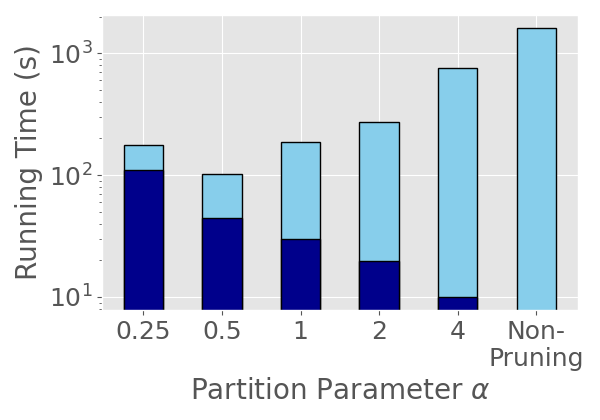}
        \vskip -6pt
        \caption{Sampling rate of 40\%}
    \end{subfigure}
    \caption{Running time of pruning and final validation in SMC based verification under different partition parameters $\alpha$.}
    \label{fig:exp-verify}
\end{figure}

\fakeparagraph{Vary Partition Parameter $\alpha$}
As shown in \figref{fig:exp-verify}, pruning at all levels of $\alpha$ can reduce the running time of verification, illustrating the effectiveness of our pruning strategy.
Besides, choosing a smaller $\alpha$ results in larger number of partitions, leading to increased pruning time but reduced final validation time.
The figure indicates that in real-world data, pruning achieves the optimal performance when $\alpha=0.5$ and bring an up to 16.2× improvement in running time.

\subsection{Experiments on Multiple Data Owners}\label{sec:exp-fed}
Our \frameworkabbr can be extended to a more general scenario where the trajectory database $TD$ is distributed among multiple data owners.
The extended method follows these steps: initially, the query user employs $T_Q$ to generate $G_Q$ and broadcasts it to all the data owners; then each data owner filters their local database using $G_Q$; finally, the query user performs verification with all the data owners in parallel. We conduct the experiment on Multi-Company dataset, a real-world trajectory dataset distributed among five data owners \cite{tong2022hu}.
Sampling rates 10\% and 40\% are used in the following experiments.

\begin{figure}
    \centering
    \begin{subfigure}{0.3\textwidth}
        \centering
        \includegraphics[width=\textwidth]{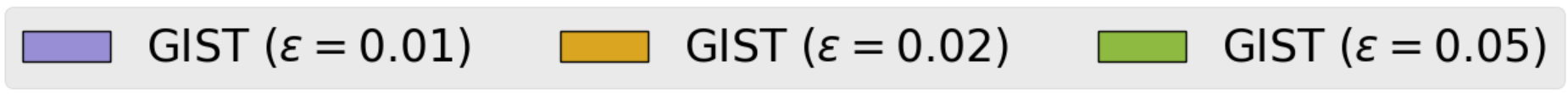}
        \caption*{}
    \end{subfigure}
    \vskip -16pt
    \begin{subfigure}{0.48\textwidth}
        \centering
        \includegraphics[width=\textwidth]{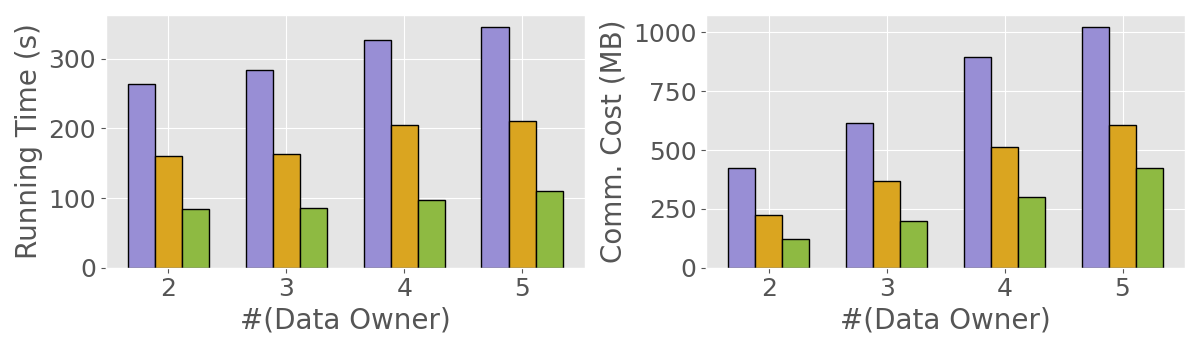}
        \vskip -6pt
        \caption{Sampling rate of 10\%}
    \end{subfigure}
    
    \begin{subfigure}{0.48\textwidth}
        \centering
        \includegraphics[width=\textwidth]{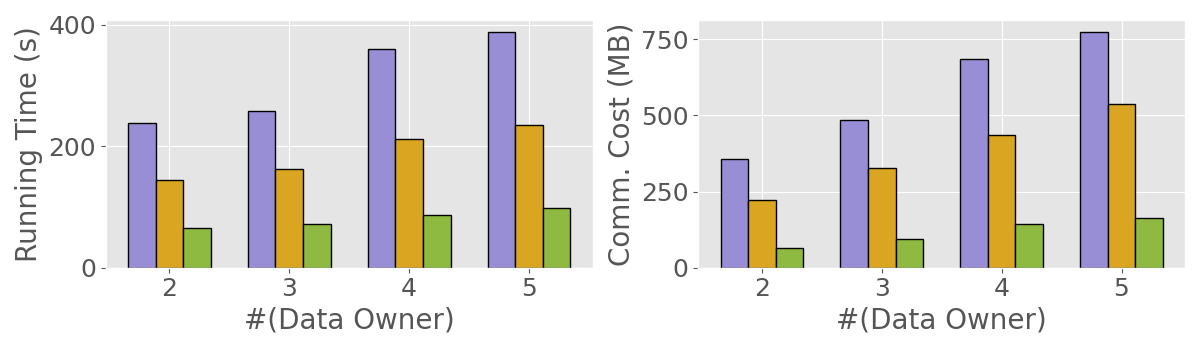}
        \vskip -6pt
        \caption{Sampling rate of 40\%}
    \end{subfigure}
    \caption{Running time and communication cost of varying the number of data owners.}
    \label{fig:exp-fed}
\end{figure}

\fakeparagraph{Vary \#(Data Owner)}
The experiment results on Multi-Company dataset are shown in \figref{fig:exp-fed}.
Overall, the total communication cost of \frameworkabbr for cross-platform data grow linearly with the number of data owners, while the running time remains relatively steady as the number of data owners increases. This is because the filtering and verification steps of \frameworkabbr can be performed in parallel.

\subsection{Summary of Major Experimental Findings}\label{sec:exp-summary}
Previous experimental results are summarized as follows:
\begin{itemize}
    \item \frameworkabbr consistently outperforms state-of-the-art solutions in terms of running time and communication cost. For instance, in the Xi'an dataset, \frameworkabbr ($\epsilon=0.01$) is 418.4× faster than STSC-ext \cite{liu2015efficient} and incurs 392.9× lower communication cost.
    \item Ablation studies show the effectiveness of our filtering and verification methods, respectively. Regarding filtering, our privacy mechanism is at least 85× more effective than the existing privacy mechanism \cite{andres2013geo,cunningham2021real,zhang2023trajectory}. For verification, pruning can bring an up to 16.2× improvement in running time.
    \item Our solution can be extended to a more general setting where $TD$ is distributed among multiple data owners. Under this setting, the running time of \frameworkabbr grow steadily as the number of data owners increases.
\end{itemize}
\section{Related Work} \label{sec:related}

We review existing studies from three categories: \textit{trajectory similarity query}, \textit{trajectory privacy preservation}, and \textit{data federation management}.

\fakeparagraph{Trajectory Similarity Query}
Various similarity measures have been proposed for trajectory data \cite{su2020survey,hu2023spatio}. 
Some measures consider spatial information only, such as DTW \cite{yi1998efficient}, ERP \cite{chen2004marriage} and EDR \cite{chen2005robust}. 
Other measures consider both spatial and temporal information, such as STLCSS \cite{vlachos2002discovering} and STED \cite{nanni2006time}. 
Recent years has witnessed the emergence of learning-based trajectory similarity measures, such as t2vec \cite{li2018deep} and ST2Vec \cite{fang2022spatio}. 
Based on different measures, existing works have designed efficient query processing solutions \cite{xie2017distributed,wang2018torch,yuan2019distributed} and trajectory analytic systems \cite{shang2018dita,fang2021dragoon,ding2018ultraman}.
However, they cannot be used for our \problemabbr problem as they usually assume no privacy protection for query users' or data owners' trajectories.

\fakeparagraph{Trajectory Privacy Preservation}
Privacy are crucial in trajectory analytics since trajectory data may disclose sensitive information like mobility patterns and personal profiles \cite{chow2011trajectory,de2013unique,jin2020trajectory}.
\textit{Geo-Indistinguishability} is widely used in protecting a user's location by injecting planar Laplacian noise, offering adaptive privacy preservation depending on the distance \cite{andres2013geo}. 
\textit{Differential privacy} has also been applied in trajectory data publish. Central differential privacy assumes all the trajectories are collected by a central server and publish perturbed trajectories \cite{xiao2015protecting,cao2019priste} or synthetic trajectories\cite{he2015dpt,jin2022frequency} with a statistical distribution similar to the original data. 
In contrast, local differential privacy \cite{cunningham2021real,zhang2023trajectory} does not rely on the central server and leverages exponential mechanism for privacy protection\cite{cunningham2021real,zhang2023trajectory}.
\textit{Spatial cloaking} aims to blur trajectories by substituting precise locations with spatial regions \cite{gruteser2003anonymous,mokbel2006new,gidofalvi2007privacy}. 
Among these methods, we compared our privacy protection mechanism BPL with \cite{andres2013geo,cunningham2021real,zhang2023trajectory}, since they are state-of-the-art solutions adaptable for filtering.

\fakeparagraph{Data Federation Management}
Data isolation has become an obstacle for cross-silo data analytics, since sharing raw data among data owners is usually prohibited due to privacy concerns.
In response to these challenges, data federation has arisen as a promising paradigm, facilitating collaborative and secure query services for data owners interested in sharing their data.
For example, SMCQL \cite{bater2017smcql}, Conclave \cite{volgushev2019conclave}, Shrinkwrap \cite{bater2018shrinkwrap} and SAQE \cite{bater2020saqe} are data management systems over relational data federation. Hu-Fu \cite{tong2022hu} is a spatial data federation system.
There are also studies on efficient processing of specific queries over a data federation (\ie ``federated computation'' as short), such as federated range aggregation \cite{shi2021efficient}, federated join aggregation \cite{wang2021secure}, and federated approximate k nearest neighbor query \cite{zhang2023approximate}.
Moreover, there are studies that combine differential privacy and SMC to design the practical protocol for real-world applications \cite{wagh2021dp,he2017composing,bater2018shrinkwrap,wang2021dp,wang2022incshrink}. These works speed up the secure queries drastically by leveraging differential privacy to remove unnecessary dummy data/operations without sacrificing too much on the privacy.
In comparison, our \problemabbr problem differs from these studies in both data type and query type.

\section{Conclusion} \label{sec:conclusion}

In this paper, we study the problem of \problem (\problemabbr) and introduce a framework called \textbf{\framework (\frameworkabbr)}. We design a novel paradigm for publishing the query trajectory at a grid level and establish the bound for the grid size under specified privacy parameters. Besides, we devise a data partition scheme along with a reference trajectory based pruning strategy to further improve efficiency. Finally, experiments show that our method is significantly faster and takes up to 3 orders of magnitude lower communication cost than the state-of-the-arts.

\bibliographystyle{IEEEtran}
\bibliography{ref}

\end{document}